# Formalizing common sense reasoning for scalable inconsistency-robust information integration using Direct Logic$^{TM}$ Reasoning and the Actor Model

**Carl Hewitt**

***This article is dedicated to Stanisław Jaśkowski, John McCarthy, Marvin Minsky and Ludwig Wittgenstein.***

## Abstract

People use common sense in their interactions with large information systems. This common sense needs to be formalized so that it can be used by computer systems. Unfortunately, previous formalizations have been inadequate. For example, classical logic is not safe for use with pervasively inconsistent information. The goal is to develop a standard foundation for reasoning in large-scale Internet applications (including sense making for natural language).

Inconsistency Robust Direct Logic is a minimal fix to Classical Logic without the rule of Classical Proof by Contradiction

$$(\Psi \vdash (\Phi \land \neg\Phi)) \vdash \neg\Psi$$

*Addition of the above transforms Inconsistency Robust Direct Logic into Classical Logic*. Inconsistency Robust Direct Logic makes the following contributions over previous work:

- *Direct* Inference[1]
- *Direct* Argumentation (argumentation directly expressed)
- *Inconsistency-robust* Natural Deduction that doesn't require artifices such as indices (labels) on propositions or restrictions on reiteration
- Intuitive inferences hold including the following:
  - Propositional Equivalences (except absorption) including Double Negation and De Morgan
  - $\lor$-Elimination (Disjunctive Syllogism), *i.e.*, $\neg\Phi, (\Phi\lor\Psi) \vdash_T \Psi$
  - Reasoning by disjunctive cases, *i.e.*, $(\Psi\lor\Phi), (\Psi \vdash_T \Theta), (\Phi \vdash_T \Omega) \vdash_T \Theta\lor\Omega$
  - Contrapositive for implication *i.e.*, $\Psi\Rightarrow\Phi$ if and only if $\neg\Phi\Rightarrow\neg\Psi$
  - Soundness, *i.e.*, $\vdash_T ((\vdash_T \Psi) \Rightarrow \Psi)$
  - Inconsistency Robust Proof by Contradiction, *i.e.*, $\vdash_T (\Psi\Rightarrow (\Phi\land\neg\Phi))\Rightarrow\neg\Psi$

A fundamental goal of Inconsistency Robust Direct Logic is to effectively reason about large amounts of pervasively inconsistent information using computer information systems.

Jaśkowski [1948] stated the following initial goal:

*To find a system* [for inconsistency robust inference] *… which:*

1) *when applied to contradictory* [information] *would not always entail overcompleteness* [i.e. infer every proposition]
2) *would be rich enough for practical inference*
3) *would have an intuitive justification*

According to Feferman [2008]: *So far as I know, it has not been determined whether such* [inconsistency robust] *logics account for "sustained ordinary reasoning", not only in everyday discourse but also in mathematics and the sciences.* Direct Logic is put forward as an improvement over classical logic with respect to Feferman's desideratum above using the following:

- **Inconsistency Robust Direct Logic** for pervasively inconsistent theories of practice[i]
- **Classical Direct Logic** for use of consistent mathematical theories in inconsistency robust theories

Direct Logic is an improvement over classical logic with respect to Feferman's desideratum above for today's information systems that are perpetually, pervasively inconsistent. Information technology needs an all-embracing system of inconsistency-robust reasoning to support practical information integration. Having such a system is important in computer science because computers must be able to carry out all inferences (including inferences about their own inference processes) without relying on humans

**Consequently, Direct Logic is proposed as a standard to replace classical logic as a mathematical foundation for Computer Science.**

Since the global state model of computation (first formalized by Turing) is inadequate to the needs of modern large-scale Internet applications the Actor Model was developed to meet this need.

Hypothesis:[ii] **All physically possible computation can be directly implemented using Actors.**

---

[i] *e.g.*, theories for climate modeling and for modeling the human brain

[ii] This hypothesis is an update to [Church 1936] that all physically computable functions can be implemented using the lambda calculus. It is a consequence of the Actor Model that there are some computations that *cannot* be implemented in the lambda calculus.

Using, the Actor Model, this paper proves that Logic Programs are not computationally universal in that there are computations that cannot be implemented using logical inference. Consequently the Logic Program paradigm is strictly less general than the Embedding of Knowledge paradigm.

## Introduction

> *Beneath the surface of the world are the rules of science. But beneath them there is a far deeper set of rules: a matrix of pure mathematics, which explains the nature of the rules of science and how it is that we can understand them in the first place.*
> Malone [2007]

Our lives are changing: ***soon we will always be online***. People use their common sense interacting with large information systems. This common sense needs to be formalized.[i]

Large-scale Internet software systems present the following challenges:

1. **Pervasive inconsistency is the norm** and consequently classical logic infers too much, i.e., anything and everything. Inconsistencies (e.g. that can be derived from implementations, documentation, and use cases) in large software systems are pervasive and despite enormous expense have not been eliminated.
2. **Concurrency is the norm.** Logic Programs based on the inference rules of mathematical logic are not computationally universal because the message order reception indeterminate computations of concurrent programs in open systems cannot be deduced using mathematical logic from propositions about pre-existing conditions. The fact that computation is not reducible to logical inference has important practical consequences. For example, reasoning used in Information Integration cannot be implemented using logical inference [Hewitt 2008a].

This paper suggests some principles and practices formalizing common sense approaches to addressing the above issues.

## Interaction *creates* Reality[2]

> [We] *cannot think of any object apart from the possibility of its connection with other things.*
> Wittgenstein, *Tractatus*

---

[i] Eventually, computer systems need to be able to address issues like the following:

- What will be the effects of increasing greenhouse gasses?
- What is the future of mass cyber surveillance?
- What can done about the increasing prevalence of metabolic syndrome?

According to [Rovelli 2008]:

> *a pen on my table has information because it points in this or that direction. We do not need a human being, a cat, or a computer, to make use of this notion of information.*[i]

Relational physics takes the following view [Laudisa and Rovelli 2008]:

- Relational physics discards the notions of absolute state of a system and absolute properties and values of its physical quantities.
- State and physical quantities refer always to the interaction, or the relation, among multiple systems.[ii]
- Nevertheless, relational physics is a complete description of reality.[iii]

According to this view, **Interaction creates reality.**[3]

### Information is a generalization of physical information in Relational Physics

Information, as used in this article, is a generalization of the physical information of Relational Physics.[iv] Information systems participate in reality and thus are both consequence and cause. Science is a large information system that investigates and theorizes about interactions. So how does Science work?

---

[i] Rovelli added: *This* [concept of information] *is very weak; it does not require* [consideration of] *information storage, thermodynamics, complex systems, meaning, or anything of the sort. In particular:*

i. *Information can be lost dynamically* ([correlated systems can become uncorrelated])*;*
ii. [It does] *not distinguish between correlation obtained on purpose and accidental correlation;*
iii. *Most important: any physical system may contain information about another physical system.*

Also, *Information is exchanged via physical interactions.* and furthermore, *It is always possible to acquire new information about a system.*

[ii] *In place of the notion of state, which refers solely to the system,* [use] *the notion of the information that a system has about another system.*

[iii] Furthermore, according to [Rovelli 2008], *quantum mechanics indicates that the notion of a universal description of the state of the world, shared by all observers, is a concept which is physically untenable, on experimental grounds*. In this regard, [Feynman 1965] offered the following advice: *Do not keep saying to yourself, if you can possibly avoid it, "But how can it be like that?" because you will go "down the drain," into a blind alley from which nobody has yet escaped.*

[iv] Unlike physical information in Relational Physics [Rovelli 2008, page 10], this paper does *not* make the assumption that information is necessarily a discrete quantity or that it must be consistent.

According to [Law 2004, emphasis added]:

> ... ***scientific routinisation, produced with immense difficulty and at immense cost, that secures the general continued stability of natural (and social) scientific reality. Elements within*** **[**this routinisation**]** ***may be overturned… But overall and most of the time, … it is the expense*** **[**and other difficulties**]** ***of doing otherwise that allows*** **[**scientific routinisation**]** ***to achieve relative stability. So it is that a scientific reality is produced that holds together more or less***.[4]

He added that we can respond as follows:

> *That we refuse the distinction between the literal and the metaphorical (as various philosophers of science have noted, the literal is always 'dead' metaphor, a metaphor that is no longer seen as such). ... That we work allegorically.* ***That we imagine coherence without consistency.*** [emphasis added]

The coherence envisaged by Law (above) is a dynamic interactive ongoing process among humans and other objects.

## Pervasive Inconsistency is the Norm in Large Software Systems

> *"... find bugs faster than developers can fix them and each fix leads to another bug"*
> Cusumano & Selby, 1995, p. 40

The development of large software systems and the extreme dependence of our society on these systems have introduced new phenomena. These systems have pervasive inconsistencies among and within the following:[5]

- *Use cases* that express how systems can be used and tested in practice.[6]
- *Documentation* that expresses over-arching justification for systems and their technologies.[7]
- *Code* that expresses implementations of systems

Adapting a metaphor used by Popper[8] for science, the bold structure of a large software system rises, as it were, above a swamp. It is like a building erected on piles. The piles are driven down from above into the swamp, but not down to any natural or given base; and when we cease our attempts to drive our piles into a deeper layer, it is not because we have reached bedrock. We simply pause when we are satisfied that they are firm enough to carry the structure, at least for the time being. Or perhaps we do something else more pressing. Under some piles there is no rock. Also some rock does not hold.

Different communities are responsible for constructing, evolving, justifying and maintaining documentation, use cases, and code for large, software systems. In specific cases any one consideration can trump the others.

Sometimes debates over inconsistencies among the parts can become quite heated, *e.g.,* between vendors. ***In the long run, after difficult negotiations, in large software systems, use cases, documentation, and code all change to produce systems with new inconsistencies. However, no one knows what they are or where they are located!***

***A large software system is never done*** [Rosenberg 2007].[9]

With respect to *detected* contradictions in large information systems, according to [Russo, Nuseibeh, and Easterbrook 2000]:

> *The choice of an inconsistency handling strategy depends on the context and the impact it has on other aspects of the development process. Resolving the inconsistency may be as simple as adding or deleting information from a software description. However, it often relies on resolving fundamental conflicts, or taking important design decisions. In such cases, immediate resolution is not the best option, and a number of choices are available:*
>
> - ***Ignore*** *- it is sometimes the case that the effort of fixing an inconsistency is too great relative to the (low) risk that the inconsistency will have any adverse consequences. In such cases, developers may choose to ignore the existence of the inconsistency in their descriptions. Good practice dictates that such decisions should be revisited as a project progresses or as a system evolves.*
> - ***Defer*** *- this may provide developers with more time to elicit further information to facilitate resolution or to render the inconsistency unimportant. In such cases, it is important to flag the parts of the descriptions that are affected, as development will continue while the inconsistency is tolerated.*
> - ***Circumvent*** *- in some cases, what appears to be an inconsistency according to the consistency rules is not regarded as such by the software developers. This may be because the rule is wrong, or because the inconsistency represents an exception to the rule that had not been captured. In these cases, the inconsistency can be circumvented by modifying the rule, or by disabling it for a specific context.*
> - ***Ameliorate*** *- it may be more cost-effective to 'improve' a description containing inconsistencies without necessarily resolving them all. This may include adding information to the description that alleviates some adverse effects of an inconsistency and/or resolves other inconsistencies as a side effect. In such cases, amelioration can be a useful inconsistency handling strategy in that it moves the development process in a 'desirable' direction in which inconsistencies and their adverse impact are reduced.*

## Inconsistency Robustness

> *You cannot be confident about applying your calculus until you know that there are no hidden contradictions in it.*[i]
> Turing *circa* 1930. [Wittgenstein 1933-1935]
>
> *Indeed, even at this stage, I predict a time when there will be mathematical investigations of calculi containing contradictions, and people will actually be proud of having emancipated themselves from consistency.*
> Wittgenstein *circa* 1930. [Wittgenstein 1933-1935][10]

Inconsistency robustness is information system performance in the face of continually pervasive inconsistencies--- a shift from the previously dominant paradigms of *inconsistency denial* and *inconsistency elimination* attempting to sweep them under the rug.[ii]

In fact, inconsistencies are pervasive throughout our information infrastructure and they affect one another. Consequently, an interdisciplinary approach is needed.

Inconsistency robustness differs from previous paradigms based on belief revision, probability, and uncertainty as follows:

- *Belief revision*: Large information systems are continually, pervasively inconsistent and there is no way to revise them to attain consistency.
- *Probability and fuzzy logic*: In large information systems, there are typically several ways to calculate probability. Often the result is that the probability is both close to 0% and close to 100%!
- *Uncertainty*: Resolving uncertainty to determine truth is not realistic in large information systems.

There are many examples of inconsistency robustness in practice including the following:

- Our economy relies on large software systems that have tens of thousands of known inconsistencies (often called "bugs") along with tens of thousands more that have yet to be pinned down even though their symptoms are sometimes obvious.

---

[i] Turing was correct that it is unsafe to use classical logic to reason about inconsistent information. Church and Turing later proved that determining whether there are hidden inconsistencies in a mathematical theory is computationally undecidable.

[ii] Inconsistency robustness builds on previous work on inconsistency tolerance, *e.g.*, [Bertossi, Hunter and Schaub 2004; Gabbay and Hunter 1991-1992; Bėziau, Carnielli and Gabbay 2007].

- Physics has progressed for centuries in the face of numerous inconsistencies including the ongoing decades-long inconsistency between its two most fundamental theories (general relativity and quantum mechanics).
- Decision makers commonly ask for the case against as well as the case for proposed findings and action plans in corporations, governments, and judicial systems.

Inconsistency robustness stands to become a more central theme for computation. The basic argument is that because inconsistency is continually pervasive in large information systems, the issue of inconsistency robustness must be addressed!

A fundamental goal of Inconsistency Robustness is to effectively reason about large amounts of information at high degrees of abstraction:

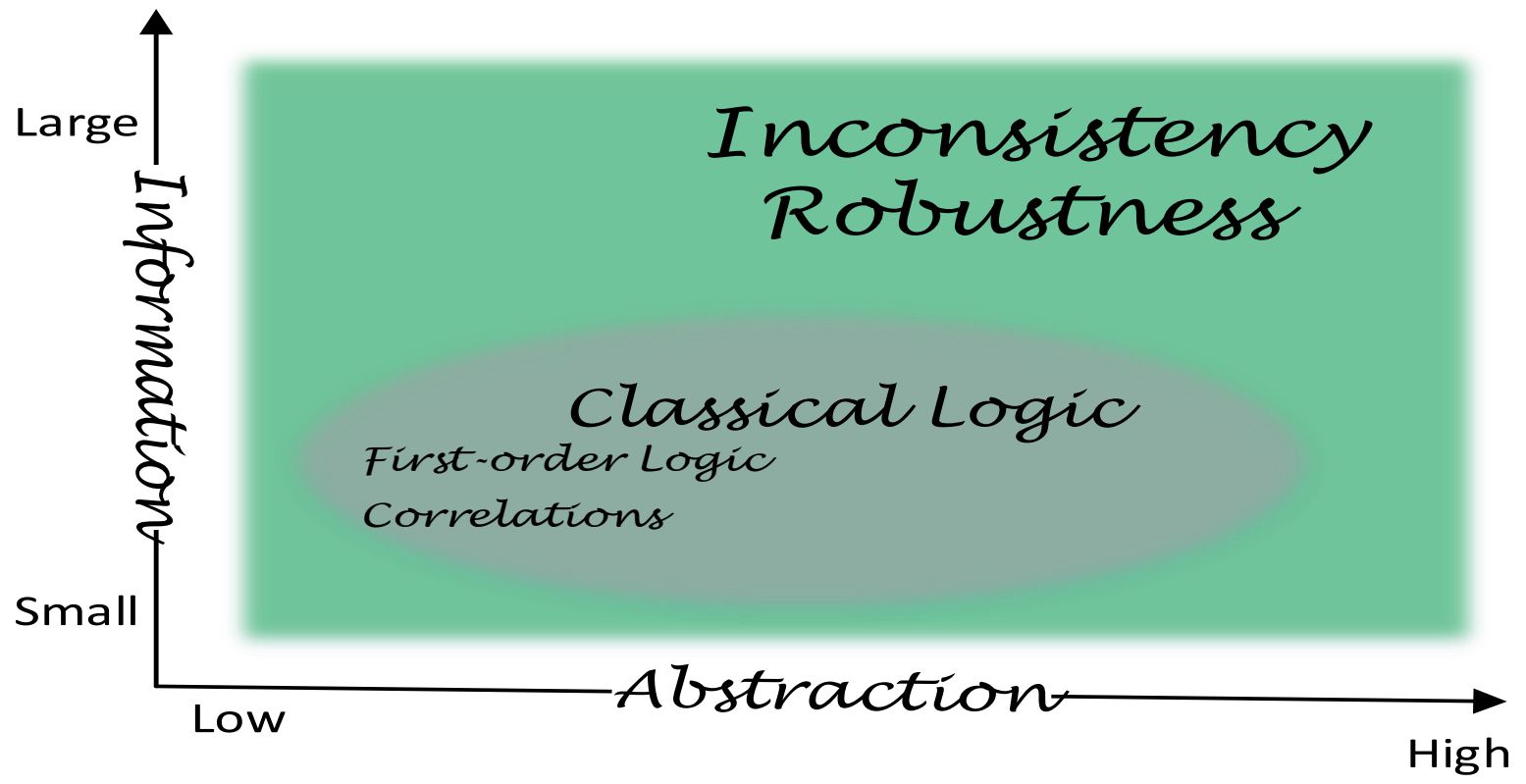

**Classical logic is safe only for theories for which there is strong evidence of consistency.**

> *A little inaccuracy sometimes saves tons of explanation.*
> Saki in "The Square Egg"

Inconsistency robust theories can be easier to develop than classical theories because perfect absence of inconsistency is not required. In case of inconsistency, there will be some propositions that can be both proved and disproved, *i.e.*, there will be arguments both for and against the propositions.

A classic case of inconsistency occurs in the novel Catch-22 [Heller 1961] which states that a person "*would be crazy to fly more missions and sane if he didn't, but if he was sane he had to fly them. If he flew them he was crazy and didn't have to; but if he didn't want to he was sane and had to. Yossarian was* *moved very deeply by the absolute simplicity of this clause of Catch-22 and let out a respectful whistle. 'That's some catch, that Catch-22,' he observed.*"

Consider the follow formalization of the above in classical logic:[i]

$\text{Policy}_1[x] \equiv \text{Sane}[x] \Rightarrow \text{Obligated}[x, \text{Fly}]$

$\text{Policy}_2[x] \equiv \text{Obligated}[x, \text{Fly}] \Rightarrow \text{Fly}[x]$

$\text{Policy}_3[x] \equiv \text{Crazy}[x] \Rightarrow \neg\text{Obligated}[x, \text{Fly}]$

$\text{Observe}_1[x] \equiv \neg\text{Obligated}[x, \text{Fly}] \wedge \neg\text{Fly}[x] \Rightarrow \text{Sane}[x]$

$\text{Observe}_2[x] \equiv \text{Fly}[x] \Rightarrow \text{Crazy}[x]$

$\text{Observe}_3[x] \equiv \text{Sane}[x] \wedge \neg\text{Obligated}[x, \text{Fly}] \Rightarrow \neg\text{Fly}[x]]$

$\text{Observe}_4 \equiv \text{Sane}[\text{Yossarian}]$

In addition, there is the following background material:

$\text{Background}_2 \equiv \neg\text{Obligated}[\text{Moon}, \text{Fly}]$

*Using classical logic,* the following rather surprising conclusion can be inferred:

**Fly[Moon]**

*i.e.,* the moon flies an aircraft!

Classical logic is not safe for theories not know to be consistent.[ii]

## Inconsistency robustness facilitates formalization

Inconsistency Robust Direct logic facilitates common sense reasoning by formalizing inconsistency robust inference.[iii]

---

[i] This is a very simple example of how classical logic can infer absurd conclusions from inconsistent information. More generally, classical inferences using inconsistent information can be arbitrarily convoluted and there is no practical way to test if inconsistent information has been used in a derivation.

[ii] It turns out that there is a hidden inconsistency in the theory *Catch22:*

$\text{Inference}_1 \equiv \ \vdash_{\mathit{Catch22}} \text{Fly}[\text{Yossarian}]$

$\text{Inference}_2 \equiv \ \vdash_{\mathit{Catch22}} \neg\text{Fly}[\text{Yossarian}]$

Thus there is an inconsistency in the theory *Catch22* concerning whether Yossarian flies.

[iii] According to [Minsky 1974]:

*The consistency that* [classical] *logic absolutely demands is not otherwise usually available – and probably not even desirable! – because consistent systems are likely to be too "weak".*

In Direct Logic, the above can be formulated using a very strong form of implication in Inconsistency Robust Direct Logic as follows in the theory *Catch22*:[11]

$\text{Policy}_1[x] \equiv\ \vdash_{\mathit{Catch22}} \text{Sane}[x] \Rightarrow \text{Obligated}[x, \text{Fly}]$

$\text{Policy}_2[x] \equiv\ \vdash_{\mathit{Catch22}} \text{Obligated}[x, \text{Fly}] \Rightarrow \text{Fly}[x]$

$\text{Policy}_3[x] \equiv\ \vdash_{\mathit{Catch22}} \text{Crazy}[x] \Rightarrow \neg\text{Obligated}[x, \text{Fly}]$

$\text{Observe}_1[x] \equiv\ \vdash_{\mathit{Catch22}} \neg\text{Obligated}[x, \text{Fly}] \wedge \neg\text{Fly}[x] \Rightarrow \text{Sane}[x]$

$\text{Observe}_2[x] \equiv\ \vdash_{\mathit{Catch22}} \text{Fly}[x] \Rightarrow \text{Crazy}[x]$

$\text{Observe3}[x] \equiv\ \vdash_{\mathit{Catch22}} \text{Sane}[x] \wedge \neg\text{Obligated}[x, \text{Fly}] \Rightarrow \neg\text{Fly}[x]]$

$\text{Observe4} \equiv\ \vdash_{\mathit{Catch22}} \text{Sane}[\text{Yossarian}]$

$\text{Background2} \equiv\ \vdash_{\mathit{Catch22}} \neg\text{Obligated}[\text{Moon}, \text{Fly}]$

Unlike Classical Logic, in Direct Logic:

$\nvdash_{\mathit{Catch22}} \text{Fly}[\text{Moon}]$

It turns out that the following can be inferred:[12]

$\vdash_{\mathit{Catch22}} \text{Fly}[\text{Yossarian}]$

$\vdash_{\mathit{Catch22}} \neg\text{Fly}[\text{Yossarian}]$

However, instead of being able to infer everything[i], once the above contradiction been noticed, question answering can be improved using the "**but**" construct of Inconsistency Robust Direct Logic as follows:

$\vdash_{\mathit{Catch22}} \text{Fly}[\text{Yossarian}]$ **but** $\vdash_{\mathit{Catch22}} \neg\text{Fly}[\text{Yossarian}]$

$\vdash_{\mathit{Catch22}} \neg\text{Fly}[\text{Yossarian}]$ **but** $\vdash_{\mathit{Catch22}} \text{Fly}[\text{Yossarian}]$

## Contradictions can facilitate Argumentation

> [I] *emphasize that contradictions are not always an entirely bad thing. I think we have all found in our googling that it is often better to find contradictory information on a search topic rather than finding no information at all. I explore some of the various reasons this may arise, which include finding that there is at least active interest in the topic, appraising the credentials of the informants, counting their relative number, assessing their arguments, trying to reproduce their experimental results, discovering their authoritative sources, etc.*
> [Dunn 2014]

[i] which is the case in classical logic from a contradiction

Using Direct Logic, various arguments can be made in $\mathcal{Catch22}$. For example:

$$\text{Sane}[x] \vdash^{\text{Argument1}}_{Catch22} \text{Crazy}[x]$$

*i.e.* "The sane ones are thereby crazy because they fly."

$$\text{Crazy}[x], \neg\text{Fly}[x] \vdash^{\text{Argument2}}_{Catch22} \text{Sane}[x]$$

*i.e.* "The crazy ones who don't fly are thereby sane."

However, neither of the above arguments is absolute because there might be arguments against the above arguments. Also, the following axiom can be added to the mix:

$$\text{Observe}_5[x] \equiv \vdash_{\mathcal{Catch22}} \text{Crazy}[x] \Rightarrow \neg\text{Sane}[x]]$$

Once, the above axiom is added we have:

$\vdash_{\mathcal{Catch22}}$ Fly[Yossarian] **but** $\vdash_{\mathcal{Catch22}}$ ¬Sane[Yossarian]

although Sane[Yossarian] is used in the argument for Fly[Yossarian].

The theory $\mathcal{Catch22}$ illustrates the following points:

- *Inconsistency robustness facilitates theory development because a single inconsistency is not disastrous.*
- *Even though the theory $\mathcal{Catch22}$ is inconsistent, it is not meaningless.*
- *Queries can be given sensible answers in the presence of inconsistent information.*

## Inconsistent probabilities

> *You can use all the quantitative data you can get, but you still have to distrust it and use your own intelligence and judgment.*
> Alvin Toffler
>
> *it would be better to ... eschew all talk of probability in favor of talk about correlation.*
> N. David Mermin [1998]

Inconsistency is built into the very foundations of probability theory:[13]

- $\mathbb{P}\text{PresentMoment} \cong 0$
  Because of cumulative contingencies to get here.[i]
- $\mathbb{P}\text{PresentMoment} \cong 1$
  Because it's reality.

The above problem is not easily fixed because of the following:

---

[i] For example, suppose that we have just flipped a coin a large number of times producing a long sequence of heads and tails. The exact sequence that has been produced is extremely unlikely.

- Indeterminacies are omnipresent/
- Interdependencies[14] are pervasive thereby calling to question probabilistic calculations that assume independence.

The above points about the perils of correlation were largely missed in [Anderson 2008]. which stated

> *"Correlation is enough." **We can stop looking for models. We can analyze the data without hypotheses about what it might show**. We can throw the numbers into the biggest computing clusters the world has ever seen and let statistical algorithms find patterns where science cannot.* (emphasis added)

Of course, Anderson missed the whole point that causality is about ***affecting*** correlations through interaction. Statistical algorithms can always find meaningless correlations. Models (*i.e.* theories) are used to create interventions to test which correlations are causal.

**Theorem.** $(\Psi \vdash \Phi) \Rightarrow \mathbb{P}\Psi \leq \mathbb{P}\Phi$

Proof: Suppose $\Psi \vdash \Phi$.

$$1 \cong^{i} \mathbb{P}\Phi|\Psi \equiv \frac{\mathbb{P}\Phi\wedge\Psi}{\mathbb{P}\Psi}$$

$$\mathbb{P}\Psi \cong \mathbb{P}\Phi\wedge\Psi \leq \mathbb{P}\Phi$$

Thus probabilities for the theory Catch22 obey the following:

**P1.** $\vdash_{Catch22}$ $\mathbb{P}$Sane[x] $\leq$ $\mathbb{P}$Obligated[x, Fly]

**P2.** $\vdash_{Catch22}$ $\mathbb{P}$Obligated[x, Fly] $\leq$ $\mathbb{P}$Fly[x]

**P3.** $\vdash_{Catch22}$ $\mathbb{P}$Crazy[x] $\leq$ $\mathbb{P}\neg$Obligated[x, Fly] ]

**S1.** $\vdash_{Catch22}$ $\mathbb{P}\neg$Obligated[x, Fly] $\wedge\neg$Fly[x] $\leq$ $\mathbb{P}$Sane[x]

**S2.** $\vdash_{Catch22}$ $\mathbb{P}$Fly[x] $\leq$ $\mathbb{P}$Crazy[x]

**S3.** $\vdash_{Catch22}$ $\mathbb{P}$Sane[x]$\wedge\neg$Obligated[x, Fly] $\leq$ $\mathbb{P}\neg$Fly[x]

**S4.** $\vdash_{Catch22}$ $\mathbb{P}$Sane[Yossarian] $\cong$ 1

Consequently, the following inferences hold

**I1.** $\vdash_{Catch22}$ 1 $\cong$ $\mathbb{P}$Obligated[Yossarian, Fly] ⓘ *using* **P1** *and* **S4**

**I2.** $\vdash_{Catch22}$ 1 $\cong$ $\mathbb{P}$Fly[Yossarian] ⓘ *using* **P2** and **I1**

**I3.** $\vdash_{Catch22}$ 1 $\cong$ $\mathbb{P}$Crazy[Yossarian] ⓘ *using* **S2** *and* **I2**

**I4.** $\vdash_{Catch22}$ 1 $\lesssim$ $\mathbb{P}\neg$Obligated[Yossarian, Fly] ⓘ *using* **P3** *and* **I3**

**I5.** $\vdash_{Catch22}$ $\mathbb{P}\neg$Fly[Yossarian] $\cong$ 0 ⓘ *using* **I4** *and* **S3**

**I6.** $\vdash_{Catch22}$ $\mathbb{P}$Fly[Yossarian] $\cong$ 1 ⓘ *reformulation of* **I5**

---

[i] This conclusion is not accepted by all. See [Lewis 1976].

Thus there is an inconsistency in Catch22 in that both of the following hold in the above:

**I2.** $\vdash_{Catch22}$ **ℙFly[Yossarian] ≅ 1**

**I6.** $\vdash_{Catch22}$ **ℙFly[Yossarian] ≅ 0**

Inconsistent probabilities are potentially a much more serious problem than logical inconsistencies because they have unfortunate consequences like the following: $\vdash_{Catch22} 1 \cong 0$.[15]

In addition to inconsistency non-robustness, probability models are limited by the following:

- ✗ Limited expressiveness (avoidance of non-numerical reasoning)
- ✗ Limited scalability
- ✗ Fragile independence assumptions
- ✗ Markovian ahistoricity
- ✗ Bayes rule (very conservative) versus general reasoning
- ✗ Contrafactuals (contra scientific knowledge)

Nevertheless, probabilities have important uses in physics, *e.g.* quantum systems.

However, statistical reasoning is enormously important in practice including the following:

- Aggregation and Correlation
- Interpolation and Extrapolation
- Classification and Simulation

## Circular information

How can inconsistencies such as the one above be understood?
Assigning truth values to propositions is an attempt to characterize whether or not a proposition holds in a theory. Of course, this cannot be done consistently if the theory is inconsistent. Likewise, assigning probabilities to propositions is an attempt to characterize the likelihood that a proposition holds in a theory. Similar to assigning truth values, assigning probabilities cannot be done consistently if the theory is inconsistent.

The process of theory development can generate circularities that are an underlying source of inconsistency:

> *Mol shows that clinical diagnoses often depend on collective and statistically generated norms. What counts as a 'normal' haemoglobin level in blood is a function of measurements of a whole population. She is saying, then, that* ***individual diagnoses include collective norms though they cannot be reduced to these*** *(Mol and Berg 1994). At the same time,* *however, the collective norms depend on a sample of clinical measurements which may be influenced by assumptions about the distribution of anaemia—though it is not, of course, reducible to any individual measurement. The lesson is that* ***the individual is included in the collective, and the collective is included in the individual—but neither is reducible to the other.***[16]

## Classical logic is unsafe for use with potentially inconsistent information

> *Irony is about contradictions that do not resolve into larger wholes even dialectically, about the tension of holding incompatible things together because all are necessary and true.*
> Haraway [1991]

An important limitation of classical logic[i] for inconsistent information is that it supports the principle that from an inconsistency anything and everything can be inferred, *e.g. "The moon is made of green cheese."*

For convenience, I have given the above principle the name IGOR[17] for **I**nconsistency in **G**arbage **O**ut **R**edux. IGOR can be formalized as follows in which a contradiction about a proposition **Ω** infers any proposition **Θ**,[ii] *i.e.*, $\mathbf{\Omega, \neg\, \Omega \vdash \Theta}$**.**

**Of course, IGOR *cannot* be part of Inconsistency Robust Direct Logic because it allows *every* proposition to be inferred from a contradiction.**

The IGOR principle of classical logic may not seem very intuitive! So why is it included in classical logic?

- **Classical Proof by Contradiction:** $(\Psi \vdash \Phi, \neg\Phi) \Rightarrow (\vdash \neg\Psi)$, which can be justified in classical logic on the grounds that if **Ψ** infers a contradiction in a consistent theory then **Ψ** must be false. In an inconsistent theory. Classical Proof by Contradiction leads to explosion by the following derivation in classical logic by a which a contradiction about P infers any proposition **Θ**:
  $$\underline{P, \neg P} \vdash \neg\underline{\Theta \vdash P, \neg P} \vdash (\neg\neg\Theta) \vdash \Theta$$
- **Classical Contrapositive for Inference:** $(\Psi \vdash \Phi) \Rightarrow (\neg\Phi \vdash \neg\Psi)$, which can be justified in classical logic on the grounds that if $\Psi \vdash \Phi$, then if **Φ** is false then **Ψ** must be false. In an inconsistent theory. Classical

---

[i] A very similar limitation holds for intuitionistic logic.

[ii] Using the symbol ⊢ to mean "infers in classical mathematical logic." The symbol was first published in [Frege 1879].

Contrapositive for Inference leads to explosion by the following derivation in classical logic by a which a contradiction about P (*i.e.,* $\vdash$ P, ¬P **)** infers any proposition **Θ** by the following proof:

Since $\vdash$ P, ¬**Θ** $\vdash$ P by monotonicity. Therefore ¬P $\vdash$ **Θ** by Classical Contrapositive for Inference. Consequently P, ¬P $\vdash$ **Θ.**

- **Classical Extraneous ∨ Introduction**:[18] Ψ $\vdash$ (Ψ∨Φ), which in classical logic says that if Ψ is true then Ψ∨Φ is true regardless of whether Φ is true.[19] In an inconsistent theory, Extraneous ∨ introduction leads to explosion via the following derivation in classical logic in which a contraction about P infers any proposition Θ:

  <u>P,¬P</u> $\vdash$ <u>(P∨Θ),¬P</u> $\vdash$ Θ
- **Classical Excluded Middle:** $\vdash$ (Ψ∨¬Ψ), which in classical logic says that Ψ∨¬Ψ is true regardless of whether Ψ is true. *Excluded Middle* is the principle of Classical Logic that for every proposition X the following holds: ExcludedMiddle[X] ≡ X∨¬X

  However, Excluded Middle is not suitable for inconsistency-robust logic because it is equivalent[i] to saying that there are no inconsistencies, *i.e.,* for every proposition X,

  Noncontradiction[X] ≡ ¬(X∧¬X)

  Using propositional equivalences, note that

  ExcludedMiddle[Φ∨Ψ] ⇔ (Ψ∨¬Ψ∨Φ)∧(Φ∨¬Φ∨Ψ)

  Consequently, ExcludedMiddle[Φ∨Ψ]⇒(Ψ∨¬Ψ∨Φ), which means that the principle of Excluded Middle implies Ψ∨¬Ψ∨Φ for all propositions Ψ and Φ. Thus the principle of Excluded Middle is not inconsistency robust because it implies every proposition Φ can be proved[ii] given any contradiction Ψ. [Kao 2011]

**Classical Logic is unsafe for inference using potentially inconsistent information.**[iii]

## Direct Logic

*"But if the general truths of Logic are of such a nature that when presented to the mind they at once command assent, wherein consists the difficulty of constructing the Science of Logic?"*

[Boole, 1853 pg. 3]

---

[i] using propositional equivalences

[ii] using ∨-*Elimination* , *i.e.,* ¬Φ**,** (Φ∨Ψ) $\vdash_T$ Ψ

[iii] Turing noted that classical logic can be used to make invalid inferences using inconsistent information "*without actually going through* [an explicit] *contradiction.*" [Diamond 1976] Furthermore, [Church 1935, Turing 1936] proved that it is computationally undecidable whether a mathematical theory of practice is inconsistent.

Direct Logic[20] is a framework: propositions have arguments for and against. Inference rules provide arguments that let you infer more propositions. Direct Logic is just a bookkeeping system that helps you keep track. It doesn't tell you what to do when an inconsistency is derived. But it does have the great virtue that it doesn't make the mistakes of classical logic when reasoning about inconsistent information.

The semantics of Direct Logic are based on argumentation. Arguments can be inferred for and against propositions. Furthermore, additional arguments can be inferred for and against these *arguments*, *e.g.*, supporting and counter arguments.[21]

Direct Logic must meet the following challenges:

- *Consistent* to avoid security holes
- *Powerful* so that computer systems can carry formalize all logical inferences
- *Principled* so that it can be easily learned by software engineers
- *Coherent* so that it hangs together without a lot of edge cases
- *Intuitive* so that humans can follow computer system reasoning
- *Comprehensive* to accommodate all forms of logical argumentation
- *Inconsistency Robust* to be applicable to pervasively inconsistent theories of practice with
  - Inconsistency Robust Direct Logic for logical inference about inconsistent information
  - Classical Direct Logic for mathematics used in inconsistency-robust theories

Inconsistency Robust Direct Logic is for reasoning about pervasively-inconsistent large software systems with the following goals:

- Provide a foundation for reasoning about the mutually inconsistent implementation, specifications, and use cases large software systems.
- Formalize a notion of "direct" inference for reasoning about inconsistent information
- Support "natural" deduction [Jaśkowski 1934][i] inference rules[ii]
- Support the usual propositional equivalences[iii]

---

[i] See discussion in [Pelletier 1999].

[ii] with the *exception* of the following:

- *Classical Proof by Contradiction* i.e., $(\Psi \vdash_T \neg\Phi, \Phi) \vdash_T \neg\Psi$
- *Extraneous* $\vee$ *Introduction*, i.e., $\Psi \vdash_T (\Phi\vee\Psi)$
- *Excluded Middle*, i.e., $\vdash_T (\Phi\vee\neg\Phi)$

[iii] with exception of absorption, which must be restricted to avoid IGOR

- $\vee$-*Elimination* , *i.e.*, $\neg\Phi$, $(\Phi\vee\Psi) \vdash_T \Psi$
- Reasoning by disjunctive cases,
  *i.e.*, $(\Psi\vee\Phi)$, $(\Psi \vdash_T\Theta)$, $(\Phi \vdash_T\Omega) \vdash_T\Theta\vee\Omega$
- Inconsistency Robust Proof by Contradiction, *i.e.*,
  $\vdash_T (\Psi\Rightarrow(\neg\Phi\wedge\Phi)) \Rightarrow \neg\Psi$
- Support abstraction among code, documentation, and use cases of large software systems. (See discussion below.)
- Provide increased safety in reasoning using inconsistent information.[i]

Consequently, Inconsistency Robust Direct Logic is well suited in practice for reasoning about large software systems.[ii]

**Adding just Classical Proof by Contradiction to Inconsistency Robust Direct Logic transforms it into a classical logic.**

The theories of Direct Logic are "open" in the sense of open-ended schematic axiomatic systems [Feferman 2007b]. The language of a theory can include any vocabulary in which its axioms may be applied, i.e., it is not restricted to a specific vocabulary fixed in advance (or at any other time). Indeed a theory can be an open system can receive new information at any time [Hewitt 1991, Cellucci 1992].

## In the argumentation lies the knowledge

> *You don't understand anything until you learn it more than one way.*
> [Minsky 2005][22]

Partly in reaction to Popper[iii], Lakatos [1967, §2]) calls the view below *Euclidean*:[23]

> *"Classical epistemology has for two thousand years modeled its ideal of a theory, whether scientific or mathematical, on its conception of Euclidean geometry. The ideal theory is a deductive system with an indubitable truth-injection at the top (a finite conjunction of axioms)—so that truth, flowing*

---

[i] by comparison with classical logic

[ii] In this respect, Direct Logic differs from previous inconsistency tolerant logics, which had inference rules that made them intractable for use with large software systems.

[iii] Proof by contradiction has played an important role in science (emphasized by Karl Popper [1962]) as formulated in his principle of refutation which in its most stark form is as follows:

If $\vdash_T\neg$Ob for some observation Ob, then it can be concluded that $T$ is refuted (in a theory called *Popper*), *i.e.*, $\vdash_{Popper}\neg T$

See Suppe [1977] for further discussion.

*down from the top through the safe truth-preserving channels of valid inferences, inundates the whole system."*

Since truth is out the window for inconsistent theories, we need a reformulation in terms of argumentation.

***Direct Argumentation***

**Inference in a theory $\mathcal{T}$ ($\vdash_{\mathcal{T}}$) carries chains of argument from antecedents to consequents.**

Direct Argumentation means that $\vdash_{\mathcal{T}}$ in a proposition actually means inference in the theory $\mathcal{T}$.[24] For example, together $\vdash_{\mathcal{T}}\Psi$ and $\Psi \vdash_{\mathcal{T}}\Phi$ infer $\vdash_{\mathcal{T}}\Phi$, which in Inconsistency Robust Direct Logic can be expressed as follows by *Direct Argumentation*: $\Psi, (\Psi \vdash_{\mathcal{T}}\Phi) \vdash_{\mathcal{T}}\Phi$

***Theory Dependence***

Inference in Inconsistency Robust Direct Logic is theory dependent. For example [Latour 2010]:

> *"Are these stone, clay, and wood idols true divinities*[i]*?"* [The Africans] *answered "Yes!" with utmost innocence: yes, of course, otherwise we would not have made them with our own hands*[ii]*! The Portuguese, shocked but scrupulous, not want to condemn without proof, gave the Africans one last chance: "You can't say both that you've made your own* [idols] *and that they are true divinities*[iii]*;* ***you have to choose****: it's either one or the other. Unless," they went on indignantly, "you really have no brains, and you're as oblivious to the principle of contraction*[iv] *as you are to the sin of idolatry." Stunned silence from the* [Africans] *who failed to see any contradiction.*[v]

As stated, there is no inconsistency in either the theory Africans or the theory Portuguese. But there is an inconsistency in the join of these theories, namely, Africans+Portuguese.

In general, the theories of Inconsistency Robust Direct Logic are inconsistent and therefore propositions cannot be consistently labeled with truth values.

---

[i] $\vdash_{\mathit{Africans}}$Divine[idols]

[ii] $\vdash_{\mathit{Africans}}$Fabricated[idols]

[iii] $\vdash_{\mathit{Portuguese}}$¬(Fabricated[idols] ∧ Divine[idols])

[iv] in *Africans+Portuguese*

[v] in *Africans*

## Information Invariance

> *Become a student of change. It is the only thing that will remain constant.*
> Anthony D'Angelo, The College Blue Book

Invariance[i] is a fundamental technical goal of Direct Logic.

> **Invariance:** Principles of Direct Logic are invariant as follows:
> 1. **Soundness of inference:** information is not increased by inference
> 2. **Completeness of inference:** all information that necessarily holds can be inferred

## Semantics of Direct Logic

The semantics of Direct Logic is the semantics of argumentation. Arguments can be made in favor of against propositions. And, in turn, arguments can be made in favor and against arguments. The notation $\vdash^{A}_{T}\Psi$ is used to express that A is an argument for Ψ in T.

The semantics of Direct Logic are grounded in the principle that every proposition that holds in a theory must have argument in its favor which can be expressed as follows:

> The principle **Inferences have Arguments** says that $\vdash_{T}\Psi$ if and only if there is an argument A for Ψ in T**,** *i.e.,* $\vdash^{A}_{T}\Psi$[ii]

For example, there is a controversy in biochemistry as to whether or not it has been shown that arsenic can support life with published arguments by Redfield[25] and NASA[26] to the following effect:

$$\vdash^{\text{Redfield}}_{\mathit{Biochemistry}}(\nvdash^{\text{NASA}}_{\mathit{Biochemistry}}\text{SupportsLife[Arsenic]})$$

[Rovelli 2011] has commented on this general situation:

> *There is a widely used notion that does plenty of damage: the notion of "scientifically proven". Nearly an oxymoron. The very foundation of science is to keep the door open to doubt. Precisely because we keep*

---

[i] Closely related to conservation laws in physics

[ii] There is a computational decision deterministic procedure $\text{Checker}_{T}$ running in linear time such that:

$$\forall[\text{a:}\textbf{Argument}, \text{s:}\textbf{Sentence}]\rightarrow \text{Checker}_{T}[\text{a}, \text{s}]=\text{True} \Leftrightarrow \vdash^{a}_{T}\lfloor s\rfloor_{T})$$

*questioning everything, especially our own premises, we are always ready to improve our knowledge. Therefore a good scientist is never 'certain'. Lack of certainty is precisely what makes conclusions more reliable than the conclusions of those who are certain: because the good scientist will be ready to shift to a different point of view if better elements of evidence, or novel arguments emerge. Therefore certainty is not only something of no use, but is in fact damaging, if we value reliability.*

A fanciful example of argumentation comes from the famous story "*What the Tortoise Said to Achilles*" [Carroll 1895].

Applied to example of the Tortoise in the stony, we have

$$\vdash^{\text{ProofOfZ(Axiom1, Axiom2)}}_{Achilles} Z$$ [27]

where

A ≡ "*Things that are equal to the same are equal to each other.*"

B ≡ "*The two sides of this Triangle are things that are equal to the same.*"

Z ≡ "*The two sides of this Triangle are equal to each other.*"

$\text{Axiom}_1 \equiv \ \vdash A, B$

$\text{Axiom}_2 \equiv A, B \ \vdash Z$

The above proposition fulfills the demand of the Tortoise that

*Whatever Logic is good enough to tell me is worth* ***writing down****.*

## Inference in Argumentation

*Scientist and engineers speak in the name of new allies that they have shaped and enrolled; representatives among other representatives, they add these unexpected resources to tip the balance of force in their favor.*
Latour [1987] Second Principle

"⊢ Elimination" (Chaining ) is a fundamental principle of inference: [28]

| |
|---|
| **⊢ Elimination (Chaining):** $\Psi, (\Psi \vdash_{T} \Phi) \ \vdash_{T} \Phi$ |
| ⓘ $\Phi$ *inferred in* $T$ *from* $\vdash_{T}\Psi$ *and* $\Psi \vdash_{T} \Phi$ |

SubArguments is another fundamental principle of inference:

| |
|---|
| **⊢ Introduction (SubArguments):** $(\vdash_{T \wedge \Psi} \Phi) \ \vdash_{T} (\Psi \vdash_{T} \Phi)$ |
| ⓘ *In* $T$, $\Psi$ *infers* $\Phi$ *when* $\Phi$ *is inferred in* $T \wedge \Psi$ |

***Please see the appendix "Detail of Direct Logic" for more information.***

## Mathematics Self Proves that it is Open

Mathematics proves that it is open in the sense that it can prove that its theorems cannot be provably computationally enumerated:[29]

**Theorem** $\vdash$Mathematics is Open

Proof.[i] Suppose to obtain a contradiction that it is possible to prove closure, *i.e.,* there is a provably computable total deterministic procedure Proof such that it is provable that

$$(\exists[\Psi\text{:Proposition}]\rightarrow\ \vdash^{p}\Psi\ )\ \Leftrightarrow\ \exists[\text{i}:\mathbb{N}]\rightarrow\ \text{Proof[i]=p}$$

As a consequence of the above, there is a provably total procedure ProvableComputableTotal that enumerates the provably total computable procedures that can be used in the implementation of the following procedure: Diagonal[i] ≡ (ProvableComputableTotal[i])[i]+1

However,

- ProvableComputableTotal[Diagonal] because Diagonal is implemented using provably computable total procedures.
- ¬ProvableComputableTotal[Diagonal] because Diagonal is a provably computable total procedure that differs from every other provably computable total procedure.

[Franzén 2004] argued that mathematics is inexhaustible because of inferential undecidability[ii] of closed mathematical theories. The above theorem that mathematics is open provides another independent argument for the inexhaustibility of mathematics.

## Contributions of Direct Logic

Inconsistency Robust Direct Logic aims to be a minimal fix to classical logic to meet the needs of information integration. (Addition of just the rule of Classical Proof by Contradiction by Inference, transforms Direct Logic into Classical Logic.) Direct Logic makes the following contributions over previous work:

- *Direct* Inference[30]
- *Direct* Argumentation (inference directly expressed)
- *Inconsistency Robustness*
- *Inconsistency-robust* Natural Deduction[31]
- Intuitive inferences hold including the following:
  - *Propositional equivalences*[iii]
  - *Reasoning by disjunctive cases*, i.e., $(\Psi\vee\Phi), (\Psi\vdash_{T}\Theta), (\Phi\vdash_{T}\Omega)\ \vdash_{T}\ \Theta\vee\Omega$
  - $\vee$-*Elimination*, i.e., $\neg\Phi, (\Phi\vee\Psi)\vdash_{T}\Psi$

---

[i] This argument appeared in [Church 1934] expressing concern that the argument meant that there is "*no sound basis for supposing that there is such a thing as logic*."

[ii] See section immediately below.

[iii] except absorption

- *Contrapositive for implication*: A proposition implies another if an only if negation of the latter implies negation of the former, *i.e.*, $\Psi \Rightarrow \Phi$ if and only if $\neg\Phi \Rightarrow \neg\Psi$
- *Soundness*: Inference rules are valid, *i.e.*, $\vdash_T ((\vdash_T \Psi) \Rightarrow \Psi)$
- *Inconsistency Robust Proof by Contradiction*: A hypothesis can be refuted by showing that it implies a contradiction, *i.e.,* $\vdash_T (\Phi \Rightarrow (\Psi \wedge \neg\Psi)) \Rightarrow \neg\Phi$

## Actor Model of Computation[32]

*The distinction between past, present and future is only a stubbornly persistent illusion.*
Einstein

Concurrency has now become the norm. However nondeterminism came first. See [Hewitt 2010b] for a history of models of nondeterministic computation.

### *What is Computation?*

*Any problem in computer science can be solved by introducing another level of abstraction.*
paraphrase of Alan Perlis

Turing's model of computation was intensely psychological.[33] He proposed the thesis that it included all of purely mechanical computation.[34]

Gödel declared that
It is "*absolutely impossible that anybody who understands the question* [What is computation?] *and knows Turing's definition should decide for a different concept.*"[35]

By contrast, in the Actor model [Hewitt, Bishop and Steiger 1973; Hewitt 2010b], computation is conceived as distributed in space where computational devices called Actors communicate asynchronously using addresses of Actors and the entire computation is not in any well-defined state. The behavior of an Actor is defined when it receives a message and at other times may be indeterminate.

Axioms of locality including *Organizational* and *Operational* hold as follows:
- *Organization:* The local storage of an Actor can include *addresses* only
  1. that were provided when it was created or of Actors that it has created
  2. that have been received in messages
- *Operation:* In response to a message received, an Actor can
  1 create more Actors

2 send messages[i] to addresses in the following:
- the message it has just received
- its local storage

3 for a serialized Actor, designate how to process the next message received[ii]

The Actor Model differs from its predecessors and most current models of computation in that the Actor model assumes the following:

- Concurrent execution in processing a message.
- The following are *not* required by an Actor: a thread, a mailbox, a message queue, its own operating system process, *etc.*
- Message passing has the same overhead as looping and procedure calling.

***Configurations versus Global States***

Computations are represented differently in Turing Machines and Actors:

1. *Turing Machine*: a computation can be represented as a global state that determines all information about the computation. It can be nondeterministic as to which will be the next global state, *e.g.*, in simulations where the global state can transition nondeterministically to the next state as a global clock advances in time, e.g., Simula [Dahl and Nygaard 1967].[36]
1. *Actors*: a computation can be represented as a configuration. Information about a configuration can be indeterminate.[iii]

Functions defined by lambda expressions [Church 1941] are special case Actors that never change.

That Actors which behave like mathematical functions exactly correspond with those definable in the lambda calculus provides an intuitive justification for the rules of the lambda calculus:

- *Lambda identifiers*: each identifier is bound to the address of an Actor. The rules for free and bound identifiers correspond to the Actor rules for addresses.
- *Beta reduction*: each beta reduction corresponds to an Actor receiving a message. Instead of performing substitution, an Actor receives addresses of its arguments.

---

[i] Likewise the messages sent can contain addresses only
1. that were provided when the Actor was created
2. that have been received in messagesthat are for Actors created here

[ii] An Actor that will never update its local storage can be freely replicated and cached.

[iii] For example, there can be messages in transit that will be delivered at some indefinite time.

The lambda calculus can be implemented in ActorScript as follows:

```
Actor thisIdentifier Identifier◁aType▷[ ]
                                        // thisIdentifier is bound to this identifier
  implements Expression◁aType▷ using
    eval[anEnvironment]→ anEnvironment.lookup[thisIdentifier]

Actor ProcedureCall◁aType, AnotherType▷
          [operator:([aType]↦ anotherType), operand:aType]
  implements Expression◁anotherType▷ using
    eval[anEnvironment]→
         (operator.eval[anEnvironment]).[operand.eval[environment]]

Actor Lambda◁aType, AnotherType▷
          [anIdentifier:Identifier◁aType▷, body:anotherType]
  implements Expression◁[aType]↦ anotherType▷ using
    eval[anEnvironment]→
      [anArgument:aType]→
        body.eval[Environment[anIdentifier,
                              // create a new environment with anIdentifier bound to
                                          anArgument,          // anArgument in
                                          anEnvironment]]      // anEnvironment
```

Note that in the above:

- All operations are local.
- The definition is modular in that each lambda calculus programming language construct is an Actor.
- The definition is easily extensible since it is easy to add additional programming language constructs.
- The definition is easily operationalized into efficient concurrent implementations.
- The definition easily fits into more general concurrent computational frameworks for many-core and distributed computation.

However, there are *nondeterministic* computable functions on integers that cannot be implemented using the nondeterministic lambda calculus. Furthermore, the lambda calculus can be very inefficient as illustrate by the theorem below:

**Theorem: In systems of practice[i], simulating an Actor system using a parallel implementation of a lambda expression (*i.e.* using purely functional programming) can be exponentially slower.**

The lambda calculus can express parallelism but not general concurrency (see discussion below).

***Actors generalize Turing Machines***

Actor systems can perform computations that are impossible by Turing Machines as illustrated by the following example:

> There is a bound on the size of integer that can be computed by an *always halting* nondeterministic Turing Machine starting on a blank tape.[37]

Plotkin [1976] gave an informal proof as follows:

> *Now the set of initial segments of execution sequences of a given nondeterministic program* P*, starting from a given state, will form a tree. The branching points will correspond to the choice points in the program. Since there are always only finitely many alternatives at each choice point, the branching factor of the tree is always finite.[38] That is, the tree is finitary. Now König's lemma says that if every branch of a finitary tree is finite, then so is the tree itself. In the present case this means that if every execution sequence of* P *terminates, then there are only finitely many execution sequences. So if an output set of* P *is infinite, it must contain a nonterminating computation.[39]*

The above proof is quite general and applies to the Abstract State Machine (ASM) model [Blass, Gurevich, Rosenzweig, and Rossman 2007a, 2007b; Glausch and Reisig 2006], which consequently are not really models of concurrency.

[i] Examples include climate models and medical diagnosis and treatment systems for cancer. A software system of practice typically has tens of millions of lines of code.

By contrast, the following Actor system can compute an integer of unbounded size using the ActorScript™ programming language [Hewitt 2010a]:

```
Unbounded ≡
 start[ ]→                          // a start message is implemented by
  Let aCounter ← Counter[ ],        // let aCounter be a new Counter
   Do ⓘaCounter.go[ ],     // send aCounter a go message and concurrently
    ⓘaCounter.stop[ ]
                     // return the value of sending aCounter a stop message

 Actor thisCounter Counter[ ]   // thisCounter is the name of this Actor
  count:=0                      // the variable count is initially 0
  continue:=True           // the variable continue is initially True
  stop[ ]→
     count                                    // return count
       afterward continue:=false
                  // continue is false for the next message received
  go[ ]→ continue �
              True ⦂                            // if continue is True,
                 Hole thisCounter.go[ ]       // send go[ ] to thisCounter after
                   after count := count+1     // incrementing count
              False ⦂ Void       // if continue is False, return Void
```

By the semantics of the Actor model of computation [Clinger 1981; Hewitt 2006], sending Unbounded a **start** message will result in sending an integer of unbounded size to the return address that was received with the **start** message.

The nondeterministic procedure Unbounded above can be axiomatized as follows:

$\forall[n\text{:Integer}]\rightarrow$

$\quad\exists[aRequest\text{:Request}, anInteger\text{:Integer}]\rightarrow$

$\quad\quad \text{Unbounded } \mathbf{sent}_{aRequest}\ \text{start}[\,]$

$\quad\quad\quad \Rightarrow \mathbf{Sent}_{\mathbf{Response}_{aRequest}}\ \text{returned}[anInteger] \wedge anInteger > n$

However, the above axiom does *not* compute any actual output! Instead the above axiom simply asserts the *existence* of unbounded outputs for **start** messages.

***Theorem.*** There are *nondeterministic* computable functions on integers that cannot be implemented by a nondeterministic Turing machine.

> *Proof*. The above Actor system implements a nondeterministic function[i] that cannot be implemented by a nondeterministic Turing machine.

The following arguments support unbounded nondeterminism in the Actor model [Hewitt 1985, 2006]:

- There is no bound that can be placed on how long it takes a computational circuit called an *arbiter* to settle. Arbiters are used in computers to deal with the circumstance that computer clocks operate asynchronously with input from outside, *e.g.*, keyboard input, disk access, network input, *etc.* So it could take an unbounded time for a message sent to a computer to be received and in the meantime the computer could traverse an unbounded number of states.
- Electronic mail enables unbounded nondeterminism since mail can be stored on servers indefinitely before being delivered.
- Communication links to servers on the Internet can be out of service indefinitely.

***Reception order indeterminacy***

Hewitt and Agha [1991] and other published work argued that mathematical models of concurrency did not determine particular concurrent computations as follows: The Actor Model[ii] makes use of arbitration for implementing the order in which Actors process message. Since these orders are in general indeterminate, they cannot be deduced from prior information by mathematical logic alone. Therefore mathematical logic cannot implement concurrent computation in open systems.

In concrete terms for Actor systems, typically we cannot observe the details by which the order in which an Actor processes messages has been determined. Attempting to do so affects the results. Instead of observing the internals of arbitration processes of Actor computations, we await outcomes.[40] Indeterminacy in arbiters produces indeterminacy in Actors.[iii]

---

[i] with graph {**start[ ]** ⇝ 0, **start[ ]** ⇝ 1, **start[ ]** ⇝ **2, …** }

[ii] Actors are the universal primitives of concurrent computation.

[iii] dashes are used solely to delineate crossing wires

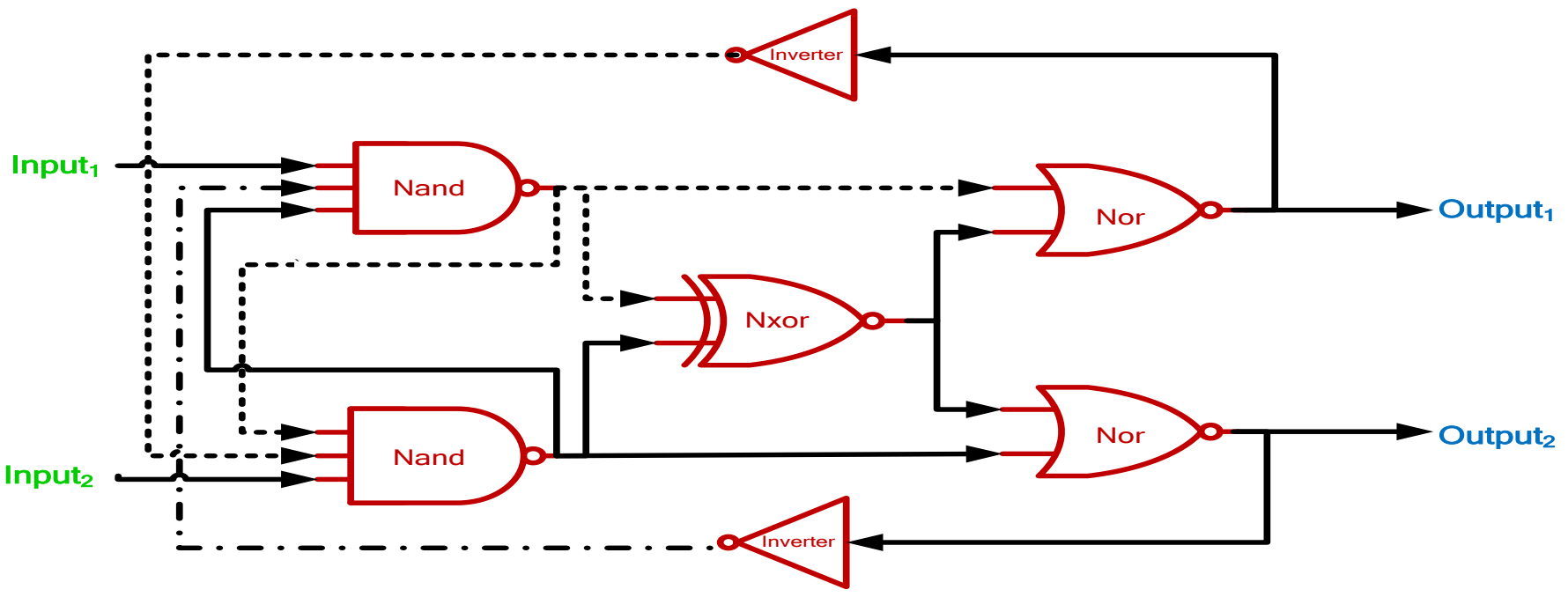

**Arbiter Concurrency Primitive**[41]

The reason that we await outcomes is that we have no realistic alternative.

### *Actor Physics*

The Actor model makes use of two fundamental orders on events [Baker and Hewitt 1977; Clinger 1981, Hewitt 2006]:

1. The *activation order* ($\rightsquigarrow$) is a fundamental order that models one event activating another (there is energy flow from an event to an event which it activates). The activation order is discrete:

   $$\forall[e_1,e_2\in\text{Events}]\rightarrow \text{Finite}[\{e\in\text{Events} \mid e_1 \rightsquigarrow e \rightsquigarrow e_2\}]$$

   There are two kinds of events involved in the activation order: reception and transmission. Reception events can activate transmission events and transmission events can activate reception events.
2. The *reception order* of a serialized Actor **x** ($\overset{\mathbf{x}}{\Rightarrow}$) models the (total) order of events in which a message is received at **x**. The reception order of each **x** is discrete:

   $$\forall[r_1,r_2\in\text{ReceptionEvents}_{\mathbf{x}}]\rightarrow \text{Finite}[\{r\in\text{ReceptionEvents}_{\mathbf{x}} \mid r_1 \overset{\mathbf{x}}{\Rightarrow} r \overset{\mathbf{x}}{\Rightarrow} r_2\}]$$

The *combined order* (denoted by $\curvearrowright$) is defined to be the transitive closure of the activation order and the reception orders of all Actors. So the following question arose in the early history of the Actor model: "*Is the combined order discrete?*" Discreteness of the combined order captures an important intuition about computation because it rules out counterintuitive computations in which an infinite number of computational events occur between two events (*à la* Zeno).

Hewitt conjectured that the discreteness of the activation order together with the discreteness of all reception orders implies that the combined order is discrete. Surprisingly [Clinger 1981; later generalized in Hewitt 2006] answered the question in the negative by giving a counterexample.

The counterexample is remarkable in that it violates the compactness theorem for 1st order logic:

Any finite set of propositions is consistent (the activation order and all reception orders are discrete) and represents a potentially physically realizable situation. But there is an infinite set of propositions that is inconsistent with the discreteness of the combined order and does not represent a physically realizable situation.

The counterexample is not a problem for Direct Logic because the compactness theorem does not hold.

The resolution of the problem is to take discreteness of the combined order as an axiom of the Actor model:

$\forall[e_1,e_2 \in \text{Events}] \rightarrow \text{Finite}[\{e \in \text{Events} \mid e_1 \curvearrowright e \curvearrowright e_2\}]$

### *Computational Representation Theorem*

*a philosophical shift in which knowledge is no longer treated primarily as referential, as a set of statements* ***about*** *reality, but as a practice that interferes with other practices. It therefore participates* ***in*** *reality.* Annemarie Mol [2002]

What does the mathematical theory of Actors have to say about the relationship between logic and computation? A closed system is defined to be one which does not communicate with the outside. Actor model theory provides the means to characterize all the possible computations of a closed system in terms of the Computational Representation Theorem [Clinger 1982; Hewitt 2006]:[42]

The denotation $\text{Denote}_S$ of a closed system S represents all the possible behaviors of S as

$$\text{Denote}_S = \lim_{i\to\infty} \text{Progression}_S{}^i$$

*where* $\text{Progression}_S$ takes a set of partial behaviors to their next stage, i.e., $\text{Progression}_S{}^i \rightarrow^{i} \text{Progression}_S{}^{i+1}$

In this way, S can be mathematically characterized in terms of all its possible behaviors (including those involving unbounded nondeterminism).[ii]

The denotations form the basis of constructively checking programs against all their possible executions,[iii]

A consequence of the Computational Representation system is that there are uncountably many different Actors.

---

[i] read as "*can evolve to*"

[ii] There are no messages in transit in $\text{Denote}_S$

[iii] a restricted form of this can be done via Model Checking in which the properties checked are limited to those that can be expressed in Linear-time Temporal Logic [Clarke, Emerson, Sifakis, *etc*. ACM 2007 Turing Award]

For example, Real▪[ ] can produce any real number[i] between 0 and 1 where

Real▪[ ] ≡ [(0 **either** 1), ∀**Postpone** Real▪[ ]]

where

- (0 **either** 1) is the nondeterministic choice of 0 or 1
- [first, ∀rest] is the sequence that begins with first and whose remainder is rest
- **Postpone** expression delays execution of expression until the value is needed.

The upshot is that **concurrent systems can be represented and characterized by logical deduction but cannot be implemented**.

Thus, the following problem arose:

> How can programming languages be rigorously defined since the proposal by Scott and Strachey [1971] to define them in terms lambda calculus failed because the lambda calculus cannot implement concurrency?

One solution is to develop a concurrent interpreter using eval messages in which eval[anEnvironment] is a message that can be sent to an expression to cause it be evaluated using the environment anEnvironment. Using such messages, modular meta-circular definitions can be concisely expressed in the Actor model for universal concurrent programming languages [Hewitt 2010a].

***Computation is not subsumed by logical deduction***

The gauntlet was officially thrown in *The Challenge of Open Systems* [Hewitt 1985] to which [Kowalski 1988b] replied in *Logic-Based Open Systems.* [Hewitt and Agha 1988] followed up in the context of the Japanese Fifth Generation Project.

Kowalski claims that "*computation could be subsumed by deduction*" [ii] His claim has been valuable in that it has motivated further research to characterize exactly which computations could be performed by Logic Programs. *However, contrary to Kowalski, computation in general is not subsumed by deduction.*

---

[i] using binary representation. See [Feferman 2012] for more on computation over the reals.

[ii] In fact, [Kowalski 1980] forcefully stated:

> *There is only one language suitable for representing information -- whether declarative or procedural -- and that is first-order predicate logic. There is only one intelligent way to process information -- and that is by applying deductive inference methods.*

### *Bounded Nondeterminism of Direct Logic*

Since it includes the nondeterministic λ calculus, direct inference, and strong induction in addition to its other inference capabilities, Direct Logic is a very powerful foundation for Logic Program languages.

But there is no Direct Logic expression that is equivalent to sending Unbounded a **start** message for the following reason:

> An expression ε will be said to always converge (written as AlwaysConverges[ε]) if and only if every reduction path terminates. *I.e.*, there is no function f such that f[0]= ε *and* ∀[i:ℕ]→⌊f[i]⌋ ↠⌊f[i+1]⌋ where the symbol ↠ is used for reduction (see the appendix of this paper on classical mathematics in Direct Logic). For example, ¬AlwaysConverges[([x]→ (0 **either** x.[x])) .[ [x]→ (0 **either** x.[x])]][i] because there is a nonterminating path.

*Theorem*: Bounded Nondeterminism of Direct Logic. If an expression in Direct Logic always converges, then there is a bound $Bound_ε$ on the number to which it can converge. *I.e.*,

∀[i:ℕ]→ (ε **AlwaysConvergesTo** n) ⇒ i≤$Bound_ε$

Consequently there is no Direct Logic program equivalent to sending Unbounded a **start** message because it has unbounded nondeterminism whereas every Direct Logic program has bounded nondeterminism.

In this way, we have proved that the Procedural Embedding of Knowledge paradigm is strictly more general than the Logic Program paradigm.

### *Computational Undecidability*

Some questions cannot be uniformly answered computationally.

The halting problem is to computationally decide whether a program halts on a given input[ii] *i.e.,* there is a total computational deterministic predicate Halt such that the following 3 properties hold for any program p and input x:

1. Halt.[p, x] $↠_1$ *True* ⇔ Converges[⌈⌊p⌋.[x]⌉]
2. Halt.[p, x] $↠_1$ *False* ⇔ ¬Converges[⌈⌊p⌋.[x]⌉]
3. Halt.[p, x] $↠_1$ *True* ∨ Halt.[p, x] $↠_1$ *False*

---

[i] Note that there are two expressions (separated by "**either**") in the bodies which provides for nondeterminism.

[ii] Adapted from [Church 1936]. Normal forms were discovered for the lambda calculus, which is the way that they "halt." [Church 1936] proved the halting problem computationally undecidable. Having done considerable work, Turing was disappointed to learn of Church's publication. The month after Church's article was published, [Turing 1936] was hurriedly submitted for publication.

[Church 1935 and later Turing 1936] published proofs that the halting problem is computationally undecidable for computable deterministic procedures.[43] In other words, there is no such procedure Halt for computable procedures.

Theorem: $\vdash \neg$ComputationallyDecidable[Halt][i]

## Classical mathematics self proves its own consistency (contra Gödel et. al.)

The following rules are fundamental to classical mathematics:

- Proof by Contradiction, *i.e.* $(\neg\Phi\Rightarrow(\Theta\wedge\neg\Theta))\vdash\Phi$, which says that a proposition can be proved showing that its negation implies a contradiction.
- Soundness, *i.e.* $(\vdash\Phi)\Rightarrow\Phi$, which says that a theorem can be used in a proof.

**Theorem:** [44] Mathematics self proves its own consistency.

*Formal Proof.* By definition, $\neg\text{Consistent}\Leftrightarrow\exists[\Psi\text{:Proposition}]\rightarrow\vdash(\Psi\wedge\neg\Psi)$. By the rule of Existential Elimination, there is some proposition $\Psi_0$ such that $\neg\text{Consistent}\Rightarrow\vdash(\Psi_0\wedge\neg\Psi_0)$ which by the rule of Soundness and transitivity of implication means $\neg\text{Consistent}\Rightarrow(\Psi_0\wedge\neg\Psi_0)$. Substituting for $\Phi$ and $\Theta$, in the rule for Proof by Contradiction, we have $(\neg\text{Consistent}\Rightarrow(\Psi_0\wedge\neg\Psi_0))\vdash\text{Consistent}$. Thus, $\vdash$Consistent.

| | |
|---|---|
| **1)** $\neg$Consistent | // hypothesis to derive a contradiction **just in this subargument** |
| **2)** $\exists[\Psi\text{:Proposition}]\rightarrow\vdash(\Psi\wedge\neg\Psi)$ | // definition of inconsistency using **1)** |
| **3)** $\vdash(\Psi_0\wedge\neg\Psi_0)$ | // rule of Existential Elimination using **2)** |
| **4)** $\Psi_0\wedge\neg\Psi_0$ | // rule of Soundness using **3)** |
| $\vdash$ Consistent | // rule of Proof by Contradiction using **1)** and **4)** |

### Natural Deduction[ii] Proof of Consistency of Mathematics

Please note the following points:

---

[i] The fact that the halting problem is computationally undecidable does not mean that proving that programs halt cannot be done in practice [Cook, Podelski, and Rybalchenko 2006].

[ii] [Jaśkowski 1934] developed Natural Deduction *cf.* [Barker-Plummer, Barwise, and Etchemendy 2011]

- The above argument formally mathematically proves that mathematics is consistent and that **it is not a premise of the theorem that mathematics is consistent.**[45]
- Classical mathematics was designed for consistent axioms and consequently the rules of classical mathematics can be used to prove consistency regardless of other axioms.[46]

The above proof means that "Mathematics is consistent" is a theorem in Classical Direct Logic. This means that the usefulness of Classical Direct Logic depends crucially on the consistency of Mathematics.[47] Good evidence for the consistency of Mathematics comes from the way that Classical Direct Logic avoids the known paradoxes. Humans have spent millennia devising paradoxes.

**The above recently developed self-proof of consistency shows that the current common understanding that Gödel proved "Mathematics cannot prove its own consistency, if it is consistent" is inaccurate.**

Long ago, Wittgenstein showed that contradiction in mathematics results from the kind of "self-referential"[i] sentence that Gödel used in his proof. However, using a typed grammar for mathematical sentences, it can be proved that the kind "self-referential" sentence that Gödel used in his proof cannot be constructed because required fixed points do not exist. In this way, consistency of mathematics is preserved without giving up power.

## Completeness versus Inferential Undecidability

"In mathematics, there is no *ignorabimus*."
Hilbert, 1902

A mathematical theory is an extension of mathematics whose proofs are computationally enumerable. For example, group theory is obtained by adding the axioms of groups along with the provision that theorems are computationally enumerable.

By definition, if T is a mathematical theory, there is a total deterministic procedure $\text{Proof}_T$ such that:

$$\vdash^{p}_{T} \psi \Leftrightarrow \exists[i:\mathbb{N}] \rightarrow \text{Proof}_T[i]=p$$

**Theorem:** If $\mathcal{T}$ is a consistent mathematical theory, there is a proposition $\psi_{\text{ChurchTuring}}$, such that both of the following hold:[i]

- $\vdash \nvdash_{\mathcal{T}} \psi_{ChurchTuring}$
- $\vdash \nvdash_{\mathcal{T}} \neg\psi_{ChurchTuring}$

Note the following important ingredients for the proof of inferential undecidability[ii] of mathematical theories:

- Closure (computational enumerability) of the theorems of a mathematical theory to carry through the proof.
- Consistency (nontriviality) to prevent everything from being provable.

Information Invariance[iii] is a fundamental technical goal of logic consisting of the following:

1. *Soundness of inference:* information is not increased by inference[iv]
2. *Completeness of inference:* all information that necessarily holds can be inferred

---

[i] Otherwise, provability in classical logic would be computationally decidable because

$$\forall[p:\textbf{Program}, x:\mathbb{N}] \rightarrow (\text{Halt}[p, x] \Leftrightarrow \vdash_{\mathcal{T}} \text{Halt}[p, x])$$

where Halt[p, x] if and only if program p halts on input x. If such a $\psi_{ChurchTuring}$ did not exist, then provability could be decided by a computable procedure $\text{Decide}_{\mathcal{T}}$:**[Sentence]↦Boolean** enumerating theorems of $\mathcal{T}$ until the proposition in question or its negation is encountered:

$$\text{Decide}_{\mathcal{T}}.[s] \rightarrow \textbf{True} \Leftrightarrow (\vdash_{\mathcal{T}} \lfloor s \rfloor) \vee \vdash_{\mathcal{T}} \neg\lfloor s \rfloor$$

Of course, $\text{Decide}_{\mathcal{T}}$ is a partial procedure and does not always converge.

[ii] sometimes called "incompleteness"

[iii] related to conservation laws in physics

[iv] *E.g.* inconsistent information does not infer nonsense.

Note that that a closed mathematical theory $T$ is inferentially undecidable with respect to $\psi_{ChurchTuring}$ does not mean "*incompleteness*" with respect to the information that can be inferred because

$$\vdash(\nvdash_T \psi_{ChurchTuring}), (\nvdash_T \neg\psi_{ChurchTuring}).$$[i]

## Information Integration

Technology now at hand can integrate all kinds of digital information for individuals, groups, and organizations so their information usefully links together.[48] Information integration needs to make use of the following information system principles:

- ***Persistence***. *Information is collected and indexed.*
- ***Concurrency****: Work proceeds interactively and concurrently, overlapping in time.*
- ***Quasi-commutativity****: Information can be used regardless of whether it initiates new work or become relevant to ongoing work.*
- ***Sponsorship****: Sponsors provide resources for computation, i.e., processing, storage, and communications.*
- ***Pluralism****: Information is heterogeneous, overlapping and often inconsistent.*
- ***Provenance****: The provenance of information is carefully tracked and recorded.*
- ***Lossless*** : Once a system has some information, then it has it thereafter.

## Opposition of Philosophers

> *By this it appears how necessary it is for nay man that aspires to true knowledge to examine the definitions of former authors; and either to correct them, where they are negligently set down, or to make them himself. For the errors of definitions multiply themselves, according as the reckoning proceeds, and lead men into absurdities, which at last they see, but cannot avoid, without reckoning anew from the beginning; in which lies the foundation of their errors...*
> [Hobbes *Leviathan*, Chapter 4]

> *Faced with the choice between changing one's mind and proving that there is no need to do so, almost everyone gets busy on the proof.* John Kenneth Galbraith [1971 pg. 50]

---

[i] by construction

A number of philosophers have opposed the results in this paper:

- Some would like to stick with just classical logic and not consider inconsistency robustness.[49]
- Some would like to stick with the first-order theories and not consider direct inference.
- Some would like to stick with just Logic Programs (*e.g.* nondeterministic Turing Machines, λ-calculus, *etc.*) and not consider concurrency.

*And some would like to have nothing to do with any of the above!*[50] However, the results in this paper (and the driving technological and economic forces behind them) tend to push towards inconsistency robustness, direct inference, and concurrency. [Hewitt 2008a]

Philosophers are now challenged as to whether they agree that

- *Inconsistency is the norm.*
- *Direct inference is the norm.*
- *Logic Programs are* ***not*** *computationally universal.*

## Scalable Information Integration

Information integration works by making connections including examples like the following:

- A statistical connection between "being in a traffic jam" and "driving in downtown Trenton between 5PM and 6PM on a weekday."
- A terminological connection between "MSR" and "Microsoft Research."
- A causal connection between "joining a group" and "being a member of the group."
- A syntactic connection between "a pin dropped" and "a dropped pin."
- A biological connection between "a dolphin" and "a mammal".
- A demographic connection between "undocumented residents of California" and "7% of the population of California."
- A geographical connection between "Leeds" and "England."
- A temporal connection between "turning on a computer" and "joining an on-line discussion."

By making these connections, iInfo™ information integration offers tremendous value for individuals, families, groups, and organizations in making more effective use of information technology.

In practice integrated information is invariably inconsistent.[51] Therefore iInfo must be able to make connections even in the face of inconsistency.[52] The business of iInfo is not to make difficult decisions like deciding the ultimate truth or probability of propositions. Instead it provides means for processing information and carefully recording its provenance including arguments (including arguments about arguments) for and against propositions.

## Work to be done

*The best way to predict the future is to invent it.* Alan Kay

There is much work to be done including the following:

### ***Invariance***

Invariance should be precisely formulated and proved. This bears on the issue of how it can be known that all the principles of Direct Logic have been discovered.

### ***Consistency***

The following conjectures for Direct Logic need to be convincingly proved:

- Consistency of Inconsistency Robust Direct Logic[i] relative to the consistency of classical mathematics. In this regard Direct Logic is consonant with Bourbaki:

  *Absence of contradiction, in mathematics as a whole or in any given branch of it, ... appears as an empirical fact, rather than as a metaphysical principle. The more a given branch has been developed, the less likely it becomes that contradictions may be met with in its farther development.*[ii]

  Thus the long historical failure to find an explosion in the methods used by Direct Logic can be considered to be strong evidence of its nontriviality.
- Constructive proof of consistency of Classical Direct Logic

### ***Inconsistency Robustness***

Inconsistency robustness of theories of Direct Logic needs to be formally defined and proved. Church remarked as follows concerning a *Foundation of Logic* that he was developing:

*Our present project is to develop the consequences of the foregoing set of postulates until a contradiction is obtained from them, or until the development has been carried so far consistently as to make it empirically probable that no contradiction can be obtained from them. And in this connection it is to be remembered that just such empirical evidence, although admittedly inconclusive, is the only existing evidence*

---

[i] *i.e.* consistency of $\vdash$

[ii] [André Weil 1949] speaking as a representative of Bourbaki

*of the freedom from contradiction of any system of mathematical logic which has a claim to adequacy.* [Church 1933][i]

Direct Logic is in a similar position except that the task is to demonstrate inconsistency robustness of inconsistent theories. This means that the exact boundaries of Inconsistency Robust Direct Logic as a minimal fix to classical logic need to be established.

***Argumentation***

Argumentation is fundamental to inconsistency robustness.

- Further work is need on fundamental principles of argumentation for large-scale information integration. See [Hewitt 2008a, 2008b].
- Tooling for Direct Logic needs to be developed to support large software systems. See [Hewitt 2008a].

***Inferential Explosion***

Inconsistencies such as the one about whether Yossarian flies are relatively *benign* in the sense that they lack significant consequences to software engineering. Other propositions (such as $\vdash_{\mathcal{T}} 1{=}0$ in a theory $\mathcal{T}$) are more *malignant* because they can be used to infer that all integers are equal to 0 using mathematical induction. To address malignant propositions, deeper investigations of argumentation using must be undertaken in which the provenance of information will play a central role. See [Hewitt 2008a].

***Robustness, Soundness, and Coherence***

Fundamental concepts such as *robustness*, *soundness*, and *coherence* need to be rigorously characterized and further developed. Inconsistency-robust reasoning beyond the inference that can be accomplished in Direct Logic needs to be developed, e.g., analogy, metaphor, discourse, debate, and collaboration.

***Evolution of Mathematics***

*In the relation between mathematics and computing science, the latter has been far many years at the receiving end, and I have often asked myself if, when, and how computing would ever be able to repay the debt.* [Dijkstra 1986]

*We argue that mathematics will become more like programming.* [Asperti, Geuvers and Natrajan 2009]

---

[i] The difference between the time that Church wrote the above and today is that the standards for adequacy have gone up dramatically. Direct Logic must be adequate to the needs of reasoning about large software systems.

Mathematical foundations are thought to be consistent by an overwhelming consensus of working professional mathematicians, e.g., mathematical theories of real numbers, integers, *etc.*

In practice, mathematical theories that are thought to be consistency by an overwhelming consensus of working mathematicians play an important supporting role for inconsistency-robust theories, *e.g.,* theories of the Liver, Diabetes, Human Behavior, *etc.*

## Conclusion

> *"The problem is that today some knowledge still feels too dangerous because our times are not so different to Cantor or Boltzmann or Gödel's time. We too feel things we thought were solid being challenged; feel our certainties slipping away. And so, as then, we still desperately want to cling to a belief in certainty. It makes us feel safe. ... Are we grown up enough to live with uncertainties or will we repeat the mistakes of the twentieth century and pledge blind allegiance to another certainty?"*
> Malone [2007]

Inconsistency robustness builds on the following principles:

- We know only a little, but it affects us *enormously*[i]
- At any point in time, much is wrong[ii] with the consensus of leading scientists but it is not known how or which parts.
- Science is never certain; it is continually (re-)made

Software engineers for large software systems often have good arguments for some proposition P and also good arguments for its negation of P. So what do large software manufacturers do? If the problem is serious, they bring it before a committee of stakeholders to try and sort it out. In many particularly difficult cases the resulting decision has been to simply live with the problem for an indefinite period. Consequently, large software systems are shipped to customers with thousands of known inconsistencies of varying severity where

- Even relatively simple subsystems can be subtly inconsistent.
- There is no practical way to test for inconsistency.
- Even though a system is inconsistent, it is not meaningless.

Inconsistency Robust Direct Logic is a minimal fix to Classical Logic without the rule of Classical Proof by Contradiction[iii], *the addition of which transforms*

[i] for better or worse
[ii] *e.g.*, misleading, inconsistent, wrong-headed, ambiguous, contra best-practices, *etc.*
[iii] *i.e.,* $(\Psi \vdash (\Phi \wedge \neg\Phi)) \vdash \neg\Psi$

*Inconsistency Robust Direct Logic into Classical Logic.* A big advantage of inconsistency robust logic is that it makes it practical for computer systems to reason about theories of practice (e.g. for macroeconomics, human history, etc.) that are pervasively inconsistent. Since software engineers have to deal with theories chock full of inconsistencies, Inconsistency Robust Direct Logic should be attractive. However, to make it relevant we need to provide them with tools that are cost effective.

Our everyday life is becoming increasingly dependent on large software systems. And these systems are becoming increasingly permeated with inconsistency and concurrency.

**As pervasively inconsistent concurrent systems become a major part of the environment in which we live, it becomes an issue of common sense to use them effectively. We will need sophisticated software systems that formalize this common sense to help people understand and apply the principles and practices suggested in this paper.**

Creating this software is not a trivial undertaking!

**There is much work to be done!**

## Acknowledgements

> *Science and politics and aesthetics, these do not inhabit different domains. Instead they interweave. Their relations intersect and resonate together in unexpected ways.*
> Law [2004 pg. 156]

Sol Feferman, Mike Genesereth, David Israel, Bill Jarrold, Ben Kuipers, Pat Langley, Vladimir Lifschitz, Frank McCabe, John McCarthy, Fanya S. Montalvo, Peter Neumann, Ray Perrault, Natarajan Shankar, Mark Stickel, Richard Waldinger, and others provided valuable feedback at seminars at Stanford, SRI, and UT Austin to an earlier version of the material in this paper. For the AAAI Spring Symposium'06, Ed Feigenbaum, Mehmet Göker, David Lavery, Doug Lenat, Dan Shapiro, and others provided valuable feedback. At MIT Henry Lieberman, Ted Selker, Gerry Sussman and the members of Common Sense Research Group made valuable comments. Reviewers for AAMAS '06 and '07, KR'06, COIN@AAMAS'06 and IJCAR'06 made suggestions for improvement.

In the logic community, Mike Dunn, Sol Feferman, Mike Genesereth, Tim Hinrichs, Mike Kassoff, John McCarthy, Chris Mortensen, Graham Priest, Dana Scott, Richard Weyhrauch and Ed Zalta provided valuable feedback

Dana Scott made helpful suggestions concerning inferential undecidability. Richard Waldinger provided extensive suggestions that resulted in better focusing a previous version of this paper and increasing its readability. Discussion with Pat Hayes and Bob Kowalski provided insight into the early history of Prolog. Communications from John McCarthy and Marvin Minsky suggested making common sense a focus. Mike Dunn collaborated on looking at the relationship of the Boolean Fragment of Inconsistency Robust Direct Logic to R-Mingle. Greg Restall pointed out that Inconsistency Robust Direct Logic does not satisfy some Relevantist principles. Gerry Allwein and Jeremy Forth made detailed comments and suggestions for improvement. Bob Kowalski and Erik Sandewall provided helpful pointers and discussion of the relationship with their work. Discussions with Ian Mason and Tim Hinrichs helped me develop Löb's theorem for Direct Logic. Scott Fahlman suggested introducing the roadmap in the introduction of the paper. At CMU, Wilfried Sieg introduced me to his very interesting work with Clinton Field on automating the search for proofs of the Gödel/Rosser inferential undecidability theorems. Also at CMU, I had productive discussions with Jeremy Avigad, Randy Bryant, John Reynolds, Katia Sycara, and Jeannette Wing. At my MIT seminar and afterwards, Marvin Minsky, Ted Selker, Gerry Sussman, and Pete Szolovits made helpful comments. Les Gasser, Mike Huhns, Victor Lesser, Pablo Noriega, Sascha Ossowski, Jaime Sichman, Munindar Singh, *etc.* provided valuable suggestions at AAMAS'07. I had a very pleasant dinner with Harvey Friedman at Chez Panisse after his 2nd Tarski lecture.

Jeremy Forth, Tim Hinrichs, Fanya S. Montalvo, and Richard Waldinger provided helpful comments and suggestions on the logically necessary inconsistencies in theories of Direct Logic. Rineke Verbrugge provided valuable comments and suggestions at MALLOW'07. Mike Genesereth and Gordon Plotkin kindly hosted my lectures at Stanford and Edinburgh, respectively, on "*The Logical Necessity of Inconsistency*". Inclusion of Cantor's diagonal argument as motivation was suggested by Jeremy Forth. John McCarthy pointed to the distinction between Logic Programs and the Logicist Programme for Artificial Intelligence. Reviewers at JAIR made useful suggestions. Mark S. Miller made important suggestions for improving the meta-circular definition of ActorScript. Comments by Michael Beeson helped make the presentation of Direct Logic more rigorous. Conversations with Jim Larson helped clarify the relationship between classical logic and the inconsistency robust logic. An anonymous referee of the Journal of Logic and Computation made a useful comment. John-Jules Meyer and Albert Visser provided helpful advice and suggestions. Comments by Mike Genesereth, Eric Kao, and Mary-Anne Williams at my Stanford Logic Group seminar "*Inference in Boolean Direct Logic is Computationally Decidable*" on 18

November 2009 greatly improved the explanation of direct inference. Discussions at my seminar "Direct Inference for Direct Logic$^{\text{TM}}$ Reasoning" at SRI hosted by Richard Waldinger on 7 January 2010 helped improve the presentation of Direct Logic. Helpful comments by Emily Bender, Richard Waldinger and Jeannette Wing improved the section on Inconsistency Robustness.

Eric Kao provided numerous helpful comments and discovered bugs in the principles of Self-refutation and Excluded Middle that were part of a previous version of Inconsistency Robust Direct Logic [Kao 2011]. Self-refutation has been replaced by Self-annihilation in the current version. Stuart Shapiro provided helpful information on why SNePS [Shapiro 2000] was based on Relevance Logic. Discussions with Dennis Allison, Eugene Miya, Vaughan Pratt and others were helpful in improving this article.

Make Travers made suggestions and comments that greatly improved the overall organization of the paper. Richard Waldinger provided guidance on classical automatic theorem provers. Illuminating conversations with Patrick Suppes provided additional ideas for improvement.

# APPENDIX 1: DETAILS OF DIRECT LOGIC

## Notation of Direct Logic

> The aims of logic should be the creation of "*a unified conceptual apparatus which would supply a common basis for the whole of human knowledge.*"
> [Tarski 1940]

In Direct Logic, unrestricted recursion is allowed in programs. For example,

- There are uncountably many Actors.[53] For example, Real▪[ ] can output any real number[i] between 0 and 1 where

  Real▪[ ] ≡ [(0 **either** 1), ⩛**Postpone** Real▪[ ]]

  where
  - (0 **either** 1) is the nondeterministic choice of 0 or 1,
  - [*first,* ⩛*rest*] is the list that begins with *first* and whose remainder is *rest*, and
  - **Postpone** *expression* delays execution of *expression* until the value is needed.
- There are uncountably many propositions (because there is a different proposition for every real number). Consequently, there are propositions that are not the abstraction of any element of a denumerable set of sentences. For example,

  $p \equiv [x \in \mathbb{R}] \rightarrow ([y \in \mathbb{R}] \rightarrow (y=x))$

  defines a different predicate p[x] for each real number x, which holds for only one real number, namely x.[ii]

It is important to distinguish between strings, sentences, and propositions. Some strings can be parsed into sentences[iii], which can be abstracted into propositions that can be asserted. Furthermore, grammar terms[iv] can be abstracted into Actors (*i.e.* objects in mathematics).

Abstraction and parsing are becoming increasingly important in software engineering. *e.g.,*

- The execution of code can be dynamically checked against its documentation. Also Web Services can be dynamically searched for and invoked on the basis of their documentation.
- Use cases can be inferred by specialization of documentation and from code by automatic test generators and by model checking.

---

[i] using binary representation.
[ii] For example (p[3])[y] holds if and only if y=3.
[iii] which are grammar tree structures
[iv] which are grammar tree structures

- Code can be generated by inference from documentation and by generalization from use cases.

**Abstraction and parsing are needed for large software systems so that that documentation, use cases, and code can mutually speak about what has been said and their relationships.**

For example:

**Proposition**

*e.g.* ∀[n:ℕ]→ ∃[m:ℕ]→ m>n

*i.e., for every* ℕ *there is a larger* ℕ

**Sentence**

*e.g.*⌈"∀[n:ℕ]→ ∃[m:ℕ]→ m>n"⌉

*i.e., the sentence that for every* ℕ *there is a larger* ℕ

**String**

*e.g.* "∀[n:ℕ]→ ∃[m:ℕ]→ m>n"

*which is a string that begins with the symbol* "∀"

In Direct Logic, a sentence is a grammar tree (analogous to the ones used by linguists). Such a grammar tree has terminals that can be constants. And there are uncountably many constants, *e.g.*, the real numbers:

The sentence $\lceil 3.14159... < 3.14159... + 1 \rceil$ is impossible to obtain by parsing a string (where 3.14159... is an Actor[i] for the transcendental real number that is the ratio of a circle's circumference to its diameter). The issue is that there is no string which when parsed is

$$\lceil 3.14159... < 3.14159... + 1 \rceil$$

Of course, because the digits of 3.14159... are computable, there is a $\mathsf{term}_1$ such that $\lfloor \mathsf{term}_1 \rfloor = 3.14159...$ that can be used to create the sentence $\lceil \mathsf{term}_1 < \mathsf{term}_1 + 1 \rceil$.

However the sentence $\lceil \mathsf{term}_1 < \mathsf{term}_1 + 1 \rceil$ is not the same as $\lceil 3.14159... < 3.14159... + 1 \rceil$ because it does not have the same vocabulary and it is a much larger sentence that has many terminals whereas $\lceil 3.14159... < 3.14159... + 1 \rceil$ has just 3 terminals:

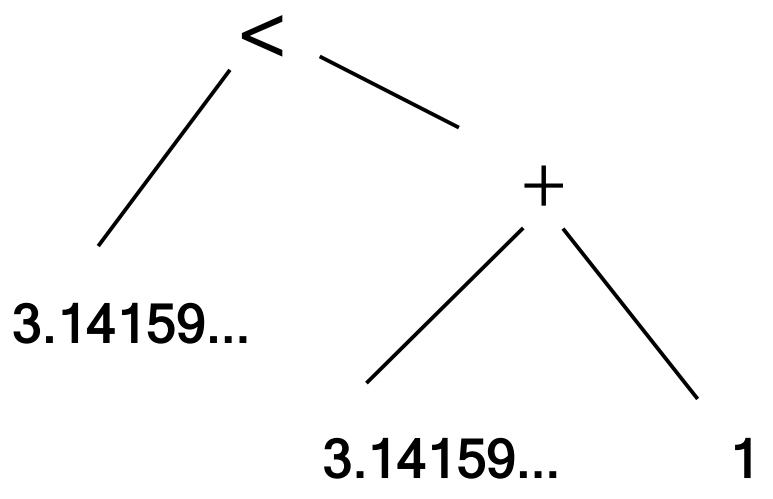

Consequently, sentences *cannot* be enumerated and there are some sentences that *cannot* be obtained by parsing strings. These arrangements exclude known paradoxes from Classical Direct Logic.[ii]

**Note: type theory of Classical Direct Logic is much stronger than constructive type theory with constructive logic[54] because Classical Direct Logic has all of the power of Classical Mathematics.**

---

[i] whose digits are incrementally computable

[ii] Please see historical appendix of this article.

Types and Propositions are defined as follows:

- ***Types***
  - **Type**, **Boolean**, $\mathbb{N}$[55], **Sentence**, **Proposition**, **Proof**, **Theory**:**Type**.
  - If $\sigma_1$, $\sigma_2$:**Type**, then $\sigma_1 \sqcup \sigma_2$, $[\sigma_1, \sigma_2]$[56], $[\sigma_1] \mapsto \sigma_2$[57], $\sigma_2{}^{\sigma_1}$[58]:**Type**.
  - If $\sigma$:**Type**, then **Term**$\lhd\sigma\rhd$[59]:**Type**.
  - If $\sigma_1$, $\sigma_2$:**Type**, $f$:$\sigma_2{}^{\sigma_1}$ and $x$:$\sigma_1$, then $f[x]$:$\sigma_2$.
  - If $\sigma_1$, $\sigma_2$:**Type**, then $\sigma_1 \sqcup \sigma_2$, $[\sigma_1] \mapsto \sigma_2$, $\sigma_2{}^{\sigma_1}$:**Type**
  - If $\sigma$:**Type**, then **Term**$\lhd\sigma\rhd$:**Type**
- ***Propositions***, *i.e.*, x:**Proposition** ⇔ x constructed by the rules below:
  - If $\sigma$:**Type**, $\Pi$:**Boolean**$^\sigma$ and $x$:$\sigma$, then $\Pi[x]$:**Proposition**.[i]
  - If $\Phi$:**Proposition**, then $\neg\Phi$:**Proposition**.
  - If $\Phi$,$\Psi$:**Proposition**, then $\Phi \wedge \Psi$, $\Phi \vee \Psi$, $\Phi \Rightarrow \Psi$, $\Phi \Leftrightarrow \Psi$:Proposition.
  - If p:**Boolean** and $\Phi$,$\Psi$:**Proposition**, then
    (p �**True**⸰ $\Phi_1$, **False**⸰ $\Phi_2$):**Proposition**.[60]
  - If $\sigma_1$,$\sigma_2$:**Type**, $x_1$:$\sigma_1$ and $x_2$:$\sigma_2$, then
    $x_1 = x_2, x_1 \in x_2, x_1 \sqsubseteq x_2, x_1 \subseteq x_2, x_1{:}x_2$:**Proposition**.
  - If T:**Theory** and $\Phi_{1 \text{ to } n}$:**Proposition**, then
    $(\Phi_1,\ldots,\Phi_k \vdash_T \Phi_{k+1},\ldots,\Phi_n)$:**Proposition** [61]
  - If T:**Theory**, p:**Proof** and $\Phi$:**Proposition**, then
    $(\vdash^{p}_{T} \Phi)$:**Proposition**[62]

---

[i] $\Pi[x] \Leftrightarrow (\Pi[x] = \text{True})$

Note that $\sigma$:**Strict**, $\Pi$:**Boolean**$^\sigma$ means that there are no fixed points for propositions.

Grammar trees (*i.e.* expressions, terms, and sentences) are defined as follows: [i]

- ***Expressions**, i.e.,* x:Expression◁σ▷ ⇔ x constructed by the rules below:
  - **True**, **False**:Constant◁Boolean▷ and **0**,**1**:Constant◁ℕ▷.
  - If σ:Type and **x**:Constant◁σ▷, then **x**:Expression◁σ▷.
  - If σ:Type and **x**:Variable◁σ▷, then **x**:Expression◁σ▷.
  - If σ,$\sigma_{1\ \text{to}\ n}$:Type, $\mathbf{x}_{1\ \text{to}\ n}$:Expression◁$\sigma_{1\ \text{to}\ n}$▷ and **y**:Expression◁σ▷, then (**Let** {$v_1 \equiv \mathbf{x}_1$ , … , $v_n \equiv \mathbf{x}_n$}[63], **y**):Expression◁σ▷ and $\mathbf{v}_{1\ \text{to}\ n}$:Variable◁$\sigma_{1\ \text{to}\ n}$▷ in **y** and in each $\mathbf{x}_{1\ \text{to}\ n}$.
  - If $e_1$, $e_2$:Expression◁Type▷, then ⌈$e_1$⊔$e_2$⌉,⌈[$e_1$, $e_2$]⌉,⌈[$e_1$]↦$e_2$⌉,⌈$e_2{}^{e_1}$⌉:Expression◁Type▷.
  - If σ:Type, $t_1$:Expression◁Boolean▷, $t_2$, $t_3$:Expression◁σ▷, then ⌈$t_1$ �True ⁏ $t_2$, False ⁏ $t_3$⌉:Expression◁σ▷.[64]
  - If $\sigma_1$,$\sigma_2$:Type, t:Expression◁$\sigma_2$▷, then ⌈[x:$\sigma_1$]→ t⌉:Expression◁[$\sigma_1$]↦$\sigma_2$▷ and **x**:Variable◁$\sigma_1$▷.[65]
  - If $\sigma_1$,$\sigma_2$:Type, **p**:Expression◁[$\sigma_1$]↦$\sigma_2$▷ and **x**:Expression◁$\sigma_1$▷, then ⌈**p**.[x] ⌉:Expression◁$\sigma_2$▷.
  - If σ:Type and **e**:Expression◁σ▷, then ⌈⌊**e**⌋⌉:Expression◁σ▷.
  - If σ:Type and **e**:Expression◁σ▷ with no free variables and **e** converges, then ⌊**e**⌋ :σ.
- ***Terms**, i.e.,* x:Term◁σ▷ ⇔ x constructed by the rules below:
  - If σ:Type and **x**:Constant◁σ▷, then **x**:Term◁σ▷.
  - If σ:Type and **x**:Variable◁σ▷, then **x**:Term◁σ▷.
  - If $t_1$, $t_2$:Term◁Type▷, then ⌈$t_1$⊔$t_2$⌉,⌈[$t_1$, $t_2$]⌉,⌈[$t_1$]↦$t_2$⌉,⌈$t_2{}^{t_1}$⌉:Term◁Type▷.
  - If σ:Type, $t_1$:Term◁Boolean▷, $t_2$,$t_3$:Term◁σ▷, then ⌈$t_1$ � True ⁏ $t_2$, False ⁏ $t_3$⌉:Term◁σ▷.
  - If $\sigma_1$,$\sigma_2$:Type, **f**:Term◁$\sigma_2{}^{\sigma_1}$▷ and **t**:Term◁$\sigma_1$▷, then ⌈f[t]⌉:Term◁$\sigma_2$▷.
  - If $\sigma_1$,$\sigma_2$:Type and **t**:Term◁$\sigma_2$▷, then ⌈[x:$\sigma_1$]→ t⌉:Term◁$\sigma_2{}^{\sigma_1}$▷ and **x**:Variable◁σ1▷ in **t**.
  - If σ:Type and **t**:Term◁σ▷, then ⌈⌊t⌋⌉:Term◁σ▷.
  - If σ:Type, **e**:Expression◁σ▷ with no free variables and **e** converges, then **e**:Constant◁σ▷.
  - If σ:Type and **t**:Term◁σ▷ with no free variables, then ⌊**t**⌋:σ.

---

[i] Because expressions are typed, fixed points do *not* exist. Parameterized mutually recursive definitions are used instead.

- *Sentences, i.e.,* x:Sentence ⇔ x constructed by the rules below:
  - If $s_1$:Sentence then, ⌈¬$s_1$⌉:Sentence.
  - If $s_1$:Sentence and $s_2$:Sentence then ⌈$s_1$∧$s_2$⌉,⌈$s_1$∨$s_2$⌉,⌈$s_1$⇨$s_2$⌉,⌈$s_1$⇔$s_2$⌉:Sentence.
  - If σ:Type, $t_1$:Term◁Boolean$^\sigma$▷ and $t_2$:Term◁σ▷, then ⌈$t_1$[$t_2$] ⌉:Sentence
  - If t:Term◁Boolean▷, $s_1$,$s_2$:Sentence, then ⌈t � True ⦂ $s_1$, False ⦂ $s_2$⌉:Sentence.[66]
  - If $\sigma_1$,$\sigma_2$:Type, $t_1$:Term◁$\sigma_1$▷ and $t_2$:Term◁$\sigma_2$▷, then ⌈$t_1$=$t_2$⌉,⌈$t_1$∈$t_2$⌉,⌈$t_1$⊑$t_2$⌉,⌈$t_1$⊂$t_2$⌉,⌈$t_1$:$t_2$⌉:Sentence.
  - If σ:Type and s:Sentence, then ⌈∀[x:σ]→ s⌉,⌈∃[x:σ]→ s⌉:Sentence and x:Variable◁σ▷ in s.
  - If T:Term◁Theory▷ and $s_{1 \text{ to } n}$:Sentence, then ⌈$s_1$, ..., $s_k$ $\vdash_T$ $s_{k+1}$, ..., $s_n$⌉:Sentence
  - If T:Term◁Theory▷, p:Term◁Proof▷ and s:Sentence, then ⌈ $\vdash^{p}_{T}$s⌉:Sentence
  - If s:Sentence, then ⌈⌊s⌋⌉:Sentence.
  - If s:Sentence with no free variables, then ⌊s⌋ :Proposition.

## Inconsistency Robust Implication

> *Whether a deductive system is Euclidean or quasi-empirical is decided by the pattern of truth value flow in the system. The system is Euclidean if the characteristic flow is the transmission of truth from the set of axioms 'downwards' to the rest of the system—logic here is an organon of proof; it is quasi-empirical if the characteristic flow is retransmission of falsity from the false basic statements 'upwards' towards the 'hypothesis'—logic here is an organon of criticism.* [Lakatos 1967]

Inconsistency-robust bi-implication is denoted by ⇔.

| |
|---|
| **Logical Equivalence:** (Ψ⇔ Φ) = (Ψ⇒Φ) ∧ (Φ⇒Ψ) |

Direct Logic has the following rules for inconsistency robust implication[i] in theory $T$:[ii]

**Reiteration:** $\vdash_T \Psi \Rightarrow \Psi$

**Exchange:** $(\vdash_T \Psi \wedge \Phi \Rightarrow \Theta) \Leftrightarrow \vdash_T \Phi \wedge \Psi \Rightarrow \Theta$

$(\vdash_T \Theta \Rightarrow \Psi \wedge \Phi) \Leftrightarrow \vdash_T \Theta \Rightarrow \Phi \wedge \Psi$

**Dropping:** $(\vdash_T \Psi \Rightarrow \Phi \wedge \Theta) \Rightarrow \vdash_T \Psi \Rightarrow \Phi$

ⓘ *an implication holds if extra conclusions are dropped*

**Accumulation:** $(\vdash_T \Psi \Rightarrow \Phi, \Psi \Rightarrow \Theta) \Rightarrow \vdash_T \Psi \Rightarrow \Phi \wedge \Theta$

$(\vdash_T \Phi \Rightarrow \Psi, \Theta \Rightarrow \Psi) \Rightarrow \vdash_T \Phi \vee \Theta \Rightarrow \Psi$

**Implication implies inference:** $(\vdash_T \Phi \Rightarrow \Psi) \Rightarrow \Phi \vdash_T \Psi$

**Transitivity:** $(\vdash_T \Psi \Rightarrow \Phi, \Phi \Rightarrow \Theta) \Rightarrow \vdash_T \Psi \Rightarrow \Theta$

ⓘ *implication in a theory is transitive*

**Contrapositive:** $(\Psi \Rightarrow \Phi) \Leftrightarrow \neg\Phi \Rightarrow \neg\Psi$

ⓘ *contrapositive holds for implication*

**Implication infers disjunction:** $(\Phi \Rightarrow \Psi) \vdash_T \Psi \vee \neg\Phi$

---

[i] denoted by ⇒. Inconsistency-robust implication is different from the much *weaker* concept of non-monotonic consequence [*e.g.* Kraus, *et. al.* 1990] which has axioms that are *not* valid for inconsistency-robust implication.

[ii] Inconsistency-robust implication is a very strong relationship. For example, monotonicity does not hold for implication although it does hold for inference. See section on Inconsistency Robust Inference below.

The ∨-rule for **Accumulation** is due to Eric Kao [private communication].

***ubstitution of Equivalent Propositions***

Logical equivalence is defined for propositions for which the usual substitution rules apply:[i]

| |
|---|
| **Substitution of equivalent propositions:** |
| $(\Psi\Leftrightarrow\Phi) \;\Rightarrow\; (\neg\Psi)\Leftrightarrow(\neg\Phi)$ |
| $(\Psi\Leftrightarrow\Phi) \;\Rightarrow\; ((\Psi\vee\Theta)\Leftrightarrow(\Phi\vee\Theta))$ |
| $(\Psi\Leftrightarrow\Phi) \;\Rightarrow\; ((\Psi\vee\Theta)\Leftrightarrow(\Phi\vee\Theta))$ |
| $(\Psi\Leftrightarrow\Phi) \;\Rightarrow\; ((\Psi\wedge\Theta)\Leftrightarrow(\Phi\wedge\Theta))$ |
| $(\Psi\Leftrightarrow\Phi) \;\Rightarrow\; ((\Psi\wedge\Theta)\Leftrightarrow(\Phi\wedge\Theta))$ |
| $(\Psi\Leftrightarrow\Phi) \;\Rightarrow\; ((\Psi\vdash_T\Theta)\Leftrightarrow(\Phi\vdash_T\Theta))$ |
| $(\Psi\Leftrightarrow\Phi) \;\Rightarrow\; ((\Theta\vdash_T\Psi)\Leftrightarrow(\Theta\vdash_T\Phi))$ |
| $(\Psi\Leftrightarrow\Phi) \;\Rightarrow\; ((\Psi\Rightarrow\Theta)\Leftrightarrow(\Phi\Rightarrow\Theta))$ |
| $(\Psi\Leftrightarrow\Phi) \;\Rightarrow\; ((\Theta\Rightarrow\Psi)\Leftrightarrow(\Theta\Rightarrow\Phi))$ |
| $(\mathbf{F}\Leftrightarrow\mathbf{G}) \;\Rightarrow\; (\forall\mathbf{F}\Leftrightarrow\forall\mathbf{G})$ |

***Propositional Equivalences***

Theorem: The following usual propositional equivalences hold:

**Self Equivalence:** $\Psi \Leftrightarrow \Psi$
**Double Negation:** $\neg\neg\Psi \Leftrightarrow \Psi$
**Idempotence of $\wedge$:** $\Psi\wedge\Psi \Leftrightarrow \Psi$
**Commutativity of $\wedge$:** $\Psi\wedge\Phi \Leftrightarrow \Phi\wedge\Psi$
**Associativity of $\wedge$:** $\Psi \wedge (\Phi\wedge\Theta) \Leftrightarrow (\Psi\wedge\Phi)\wedge \Theta$
**Distributivity of $\wedge$ over $\vee$:** $\Psi \wedge (\Phi\vee\Theta) \Leftrightarrow (\Psi\wedge\Phi) \vee (\Psi\wedge\Theta)$
**De Morgan for $\wedge$:** $\neg(\Psi\wedge\Phi) \Leftrightarrow \neg\Psi\vee\neg\Phi$
**Idempotence of $\vee$:** $\Psi\vee\Psi \Leftrightarrow \Psi$
**Commutativity of $\vee$:** $\Psi\vee\Phi \Leftrightarrow \Phi\vee\Psi$
**Associativity of $\vee$:** $\Psi \vee (\Phi\vee\Theta) \Leftrightarrow (\Psi\vee\Phi) \vee \Theta$
**Distributivity of $\vee$ over $\wedge$:** $\Psi \vee (\Phi\wedge\Theta) \Leftrightarrow (\Psi\vee\Phi)\wedge (\Psi\vee\Theta)$
**De Morgan for $\vee$:** $\neg(\Psi\vee\Phi) \Leftrightarrow \neg\Psi\wedge\neg\Phi$
**Contrapositive for $\Rightarrow$:** $(\Psi\Rightarrow\Phi) \Leftrightarrow \neg\Phi\Rightarrow\neg\Psi$

---

[i] Classical implication (denoted by ⇒) is logical implication for classical mathematics. (See the appendix on classical mathematics in Direct Logic.) Likewise classical bi-implication is denoted by ⇔.

Direct Logic has the following usual principles for equality:

$E_1=E_1$

$E_1=E_2 \Rightarrow E_2=E_1$

$(E_1=E_2 \wedge E_2=E_3) \Rightarrow E_1=E_3$

Also, the following usual propositional inferences hold:

**Absorption of ∧:** $\Psi \wedge (\Phi \vee \Psi) \vdash_T \Psi$
**Absorption of ∨:** $\Psi \vee (\Phi \wedge \Psi) \vdash_T \Psi$[67]

***Conjunction, i.e., comma***

| |
|---|
| **∧ -Elimination:** $(\vdash_T \Psi \wedge \Phi) \Rightarrow \vdash_T \Psi, \Phi$ |
| **∧ -Introduction:** $(\vdash_T \Psi, \Phi) \Rightarrow \vdash_T \Psi \wedge \Phi$ |

***Disjunction***

| |
|---|
| **∨-Elimination:**[i] $\vdash_T \neg\Psi \wedge (\Psi \vee \Phi) \Rightarrow \Phi$ |
| **∨-Introduction:** $\vdash_T \Psi \wedge \Phi \Rightarrow \Psi \vee \Phi$ |
| **Disjunctive Cases:** $\vdash_T (\Psi \vee \Phi) \wedge (\Psi \Rightarrow \Theta) \wedge (\Phi \Rightarrow \Omega) \Rightarrow \Theta \vee \Omega$ |

Theorem: ***Inconsistency Robust Resolution***[ii]

$$\vdash_T (\Psi \vee \neg\Psi) \wedge (\Psi \vee \Theta) \wedge (\Phi \vee \Omega) \Rightarrow \Theta \vee \Omega$$

Proof: Immediate from Disjunctive Cases and ∨-Elimination.

## Inconsistency Robust Inference

*Logic merely sanctions the conquests of the intuition.*
Jacques Hadamard (quoted in Kline [1972])

Inference in theory $T$ (denoted by $\vdash_T$) is characterized by the following additional axioms:[iii]

---

[i] i.e. Disjunctive Syllogism

[ii] Joint work with Eric Kao

[iii] Half of the Classical Deduction Theorem holds for Inconsistency Direct Logic. That one proposition infers another in a theory does not in general imply that the first proposition implies the second because Inconsistency Robust Implication is a *very* strong relationship.

***Soundness***

| |
|---|
| **Soundness:** $\vdash_T ((\vdash_T \Psi) \Rightarrow \Psi)$ <br> ⓘ *a proposition inferred in a theory implies the proposition in the theory* |

***Inconsistency Robust Proof by Contradiction***

| |
|---|
| **Inconsistency Robust Proof by Contradiction:** <br> $\vdash_T (\Phi \Rightarrow (\Psi \land \neg\Psi)) \Rightarrow \neg\Phi$ |

***Quantifiers***

Direct Logic makes use of functions for quantification.[68] For example following expresses commutativity for natural numbers:

$\forall[x,y:\mathbb{N}] \rightarrow x+y=y+x$

| |
|---|
| **Variable Elimination:** $\forall F \Rightarrow F[E]$ <br> ⓘ *a universally quantified variable of a statement can be instantiated with any expression* E (*taking care that none of the variables in* E *are captured*). <br> **Variable Introduction:** *Let* Z *be a new constant,* F[Z] ⇔ ∀F <br> ⓘ *inferring a statement with a universally quantified variable is equivalent to inferring the statement with a newly introduced constant substituted for the variable* <br> **Existential quantification:** $\exists F = \neg\forall\neg F$ |

## Appendix 2. Foundations of Classical Mathematics beyond Logicism

> *Mathematicians do not study objects, but the relations between objects; to them it is a matter of indifference if these objects are replaced by others, provided that the relations do not change. Matter does not engage their attention, they are interested by form alone.*
> Poincaré [1902]

This appendix presents foundations for mathematics that goes beyond logicism in that it does not attempt to reduce mathematics solely to logic, solely to types, or solely to sets in a way that encompasses all of standard mathematics including the integers, reals, analysis, geometry, *etc.*[69]

### Consistency has been the bedrock of classical mathematics.

Platonic Ideals were to be perfect, unchanging, and eternal.[70] Beginning with the Hellenistic mathematician Euclid [*circa* 300BC] in Alexandria, theories were intuitively supposed to be consistent.[71] Wilhelm Leibniz, Giuseppe Peano, George Boole, Augustus De Morgan, Richard Dedekind, Gottlob Frege, Charles Peirce, David Hilbert, *etc.* developed mathematical logic. However, a crisis occurred with the discovery of the logical paradoxes based on self-reference by Burali-Forti [1897], Cantor [1899], Russell [1903], *etc.* In response Russell [1925] stratified types, [Zermelo 1905, Fränkel 1922, Skolem 1922] stratified sets and [Tarski and Vaught 1957] stratified logical theories to limit self-reference. [Church 1935, Turing 1936] proved that closed mathematical theories are inferentially undecidable[i], *i.e.,* there are propositions which can neither be proved nor disproved. However, the bedrock of consistency remained.

This appendix present classical mathematics in Direct Logic using $\vdash$.[ii]

The following additional principles are available because $\vdash$ is thought to be consistent by an overwhelming consensus of working professional mathematicians:

**Classical Proof by Contradiction: $(\Phi \vdash \Psi, \neg\Psi)\ \vdash \neg\Phi$**
*i.e., the negation of a proposition can be inferred from inferring a contradiction*

**Classical Deduction Theorem: $(\Psi \vdash \Phi) \Leftrightarrow\ \vdash (\Psi \Rightarrow \Phi)$**
*i.e., an implication can be proved by inference*

***Inheritance from classical mathematics***

Theorems of mathematics hold in every theory:
If $\Phi$ is a proposition of mathematics, $(\vdash \Phi) \Rightarrow (\vdash_T \Phi)$

---
[i] sometimes called "incomplete"
[ii] with no subscripted inconsistency robust theory, i.e., $\vdash$ is used for classical mathematics whereas $\vdash_T$ is used for inconsistency-robust inference in theory T.

***Nondeterministic Execution***

Direct Logic makes use of the nondeterministic execution as follows:[72]

- If $\mathbf{E_1}$ and $\mathbf{E_2}$ are *expressions*, then $\mathbf{E_1} \twoheadrightarrow \mathbf{E_2}$ ($\mathbf{E_1}$ *can nondeterministically evolve to* $\mathbf{E_2}$) is a *proposition*.
- If **E** is an *expression*, then Converges[**E**] (**E** always converges) is a *proposition*.

## Foundations with both Types and Sets

Classical Direct Logic develops foundations for mathematics using *both*[i] types[ii] *and* sets[iii] encompassing all of standard mathematics including the integers, reals, analysis, geometry, *etc.*[73]

Combining types and sets as the foundation has the advantage of using the strengths of each without the limitations of trying to use just one because each can be used to make up for the limitations of the other. The key idea is compositionality, *i.e.*, composing new entities from others. Types can be composed from other types and sets can be composed from other sets.[iv]

Functions are fundamental to Computer Science. Consequently, graphs of functions and sets are fundamental collections.[74] SetFunctions◁**σ**▷ (type of set functions based on type **σ**) that can be defined inductively as follows:

$$\text{SetFunctionsOfOrder}\triangleleft\boldsymbol{\sigma}\triangleright[1] \equiv \boldsymbol{\sigma}^{\boldsymbol{\sigma}}$$

$$\text{SetFunctionsOfOrder}\triangleleft\boldsymbol{\sigma}\triangleright[n+1] \equiv (\boldsymbol{\sigma}\sqcup\text{SetFunctionsOfOrder}\triangleleft\boldsymbol{\sigma}\triangleright[n])^{\boldsymbol{\sigma}\sqcup\text{SetFunctionsOfOrder}\triangleleft\boldsymbol{\sigma}\triangleright[n]}$$

---

[i] Past attempts to reduce mathematics to logic alone, to sets alone, or to types alone have not be very successful.

[ii] According to [Scott 1967]: *"*there is only one satisfactory way of avoiding the paradoxes: namely, the use of some form of the *theory of types*... the best way to regard Zermelo's theory is as a simplification and extension of Russell's ...*simple* theory of types. Now Russell made his types *explicit* in his notation and Zermelo left them *implicit*. It is a mistake to leave something so important invisible...*"*

[iii] According to [Scott 1967]: *"*As long as an idealistic manner of speaking about abstract objects is popular in mathematics, people will speak about collections of objects, and then collections of collections of ... of collections. In other words *set theory is inevitable."* [emphasis in original]

[iv] Compositionality avoids standard foundational paradoxes. For example, Direct Logic composes propositions from others using strict types so there are no "self-referential" propositions.

Furthermore the process of constructing orders of SetFunctionsOfOrder◁σ▷ is exhaustive for SetFunctions◁σ▷:[i]

SetFunctions◁σ▷≡ $\coprod_{i:\mathbb{N}}$ **SetFunctionsOfOrder**◁σ▷[i]

Sets (along with lists) provide a convenient way to collect together elements.[75] For example, sets (of sets of sets of ...) of **σ** can be axiomatized as follows:

∀[s:**Sets**◁σ▷]→ ∃[f:**SetFunctions**◁σ▷]→ CharacteristicFunction[f, s]

where ∀[s:**Sets**◁σ▷, f:**Boolean**$^{\text{SetFunctions}◁σ▷}$]→

CharacteristicFunction[f, s]

⇔ ∀[e:σ⊔**Sets**◁σ▷]→ e∈s ⇔ f[e]=True

*i.e.* every set of type **Sets**◁σ▷ is defined by a characteristic function of **SetFunctions**◁σ▷

Note that there is no set corresponding to the *type* **Sets**◁ℕ▷ which is an example of how types extend the capabilities of sets.[76]

Although **Sets**◁ℕ▷ are well-founded[77], in general sets in Direct Logic are not well-founded. For example, consider the following definition:

InfinitelyDeep.[ ] ≡ {**Postpone** InfinitelyDeep.[ ]}[ii]

Consequently, InfinitelyDeep.[ ]∈InfinitelyDeep.[ ].

***XML***

*We speak in strings, but think in trees.*
---Nicolaas de Bruijin[78]

The base domain of Direct Logic is XML[iii]. In Direct Logic, a dog is an XML dog, *e.g.*, <Dog><Name>Fido</Name></Dog>∈Dogs⊆XML
Unlike First Order Logic, there is no unrestricted quantification in Direct Logic. So the proposition ∀d∈Dogs → Mammal[d] is about dogs in XML. *The base equality built into Direct Logic is equality for* XML, *not equality in some abstract "domain".* In this way Direct Logic does not have to take a stand on the various ways that dogs, photons, quarks and everything else can be considered "equal"!

This axiomization omits certain aspects of standard XML, *e.g.,* attributes, namespaces, *etc*.

---

[i] The closure property below is used to guarantee that there is just one model of **SetFunctions**◁ℕ▷ up to isomorphism using a unique isomorphism.

[ii] InfinitelyDeep.[ ] = {{{{{...}}}}}

[iii] Lisp was an important precursor of XML. The Atomics axiomatised below correspond roughly to atoms and the Elements to lists.

Two XML expressions are equal if and only if they are both atomic and are identical or are both elements and have the same tag and the same number of children such that the corresponding children are equal.

The following are axioms for XML:

(Atomics ∪ Elements) = XML
(Atomics ∩ Elements) = { }[79]
Tags ⊆ Atomics
∀[x]→ x∈Elements ⇔ x= <Tag(x)> $x_1 \ldots x_{\text{Length}(x)}$ </Tag(x)>
*where* $x_i$ *is the* i*th subelement of* x *and*
Tag(x) *is the tag of* x
Length(x) *is the number of subelements of* x

A set p⊆XML is defined to be *inductive* (written Inductive[p]) if and only it contains the atomics and for all elements that it contains, it also contains every element with those sub-elements:
(∀[p⊆XML; $x_1 \ldots x_n$∈p; t∈Tags]→
Inductive[p] ⇒ (Atomics ⊆ p ∧ <t> $x_1 \ldots x_n$</t>∈p)
The Strong Principle of Induction for XML is as follows:
∀[p⊆XML]→ Inductive[p] ⇒ p = XML
The reason that induction is called "*strong*" is that there are no restrictions on inductive predicates.[80]

**Natural Numbers, Real Numbers, and their Sets are Unique up to Isomorphism**

The following question arises: What mathematics have been captured in the above foundations?

**Theorem**[i] **(Categoricity of $\mathbb{N}$):**[81] $\forall[\mathbb{M}{:}\mathbf{Model}\triangleleft\mathbb{N}\triangleright]\rightarrow \mathbb{M}\approx\mathbb{N}$, *i.e.*, models of the natural numbers $\mathbb{N}$ are isomorphic by a unique isomorphism.[ii]

The following strong induction axiom[82] can be used to characterize the natural numbers ($\mathbb{N}$[83]) up to isomorphism with a unique isomorphism:

$$\forall[P{:}\mathbf{Boolean}^{\mathbb{N}}]\rightarrow\ \mathrm{Inductive}[P]\Rightarrow \forall[i{:}\mathbb{N}]\rightarrow P[i]$$

$$\textit{where}\ \forall[P{:}\mathbf{Boolean}^{\mathbb{N}}]\rightarrow$$

$$\mathrm{Inductive}[P] \Leftrightarrow P[0] \wedge \forall[i{:}\mathbb{N}]\rightarrow P[i] \Rightarrow P[i+1]$$[iii]

**Theorem**[iv] **(Categoricity of $\mathbb{R}$):**[84] $\forall[\mathbb{M}{:}\mathbf{Model}\triangleleft\mathbb{R}\triangleright]\rightarrow \mathbb{M}\approx\mathbb{R}$, *i.e.*, models of the real numbers $\mathbb{R}$ are isomorphic by a unique isomorphism.[v]

The following can be used to characterize the real numbers ($\mathbb{R}$[85]) up to isomorphism with a unique isomorphism:

$$\forall[S{:}\mathbf{Set}\triangleleft\mathbb{R}\triangleright]\rightarrow\ S\neq\{\,\} \wedge \mathrm{Bounded}[S] \Rightarrow \mathrm{HasLeastUpperBound}[S]$$

*where*

$$\mathrm{UpperBound}[b{:}\mathbb{R}, S{:}\mathbf{Set}\triangleleft\mathbb{R}\triangleright] \Leftrightarrow\ b\in S \wedge \forall[x\in S]\rightarrow x\leqq b$$

$$\mathrm{HasLeastUpperBound}[S{:}\mathbf{Set}\triangleleft\mathbb{R}\triangleright]]$$

$$\Leftrightarrow \exists[b{:}\mathbb{R}]\rightarrow \mathrm{LeastUpperBound}[b, S]$$

$$\mathrm{LeastUpperBound}[b{:}\mathbb{R}, S{:}\mathbf{Set}\triangleleft\mathbb{R}\triangleright]$$

$$\Leftrightarrow\ \mathrm{UpperBound}[b,S] \wedge \forall[x\in S]\rightarrow \mathrm{UpperBound}[x,S] \Rightarrow x\leqq b$$

**Theorem (Categoricity of Sets$\triangleleft\mathbb{N}\triangleright$):**[86]

$$\forall[\mathbb{M}{:}\mathbf{Model}\triangleleft\mathbf{Sets}\triangleleft\mathbb{N}\triangleright\triangleright]\rightarrow \mathbb{M}\approx\mathbb{S}\mathrm{ets}\triangleleft\mathbb{N}\triangleright$$

*i.e.*, models of $\mathbf{Sets}\triangleleft\mathbb{N}\triangleright$ are isomorphic by a unique isomorphism.[vi]

---

[i] [Dedekind 1888, Peano 1889]

[ii] Consequently, the type $\mathbb{N}$ is unique up to isomorphism and the type $\mathbb{R}$ is unique up to isomorphism.

[iii] which can be equivalently expressed as:

$$\forall[P{:}\mathbf{Boolean}^{\mathbb{N}}]\rightarrow\ \mathrm{Inductive}[P]\Rightarrow \forall[i{:}\mathbb{N}]\rightarrow P[i]=\mathrm{True}$$

$$\textit{where}\ \forall[P{:}\mathbf{Boolean}^{\mathbb{N}}]\rightarrow$$

$$\mathrm{Inductive}[P] \Leftrightarrow (P[0]=\mathrm{True} \wedge \forall[i{:}\mathbb{N}]\rightarrow P[i]=\mathrm{True} \Rightarrow P[i+1]=\mathrm{True})$$

[iv] [Dedekind 1888]

[v] Consequently, the type of natural numbers $\mathbb{N}$ is unique up to isomorphism and the type of reals $\mathbb{R}$ is unique up to isomorphism.

[vi] Consequently, the set of natural numbers $\mathbb{N}$ is unique up to isomorphism and is contained in the set of reals $\mathbb{R}$ that is unique up to isomorphism.

**Sets**◁ℕ▷ (which is a fundamental type of mathematics) is exactly characterized axiomatically, which is what is required for Computer Science.

Proof: By above, ∀[𝕄:**Model**◁ℕ▷]→ 𝕄≈**ℕ**, *i.e.,* models of **ℕ** are isomorphic by a unique isomorphism. Unique isomorphism of higher order sets can be proved using induction from the following closure property for SetFunctions (see above):

**SetFunctionsOfOrder**◁σ▷[n+1] ≡

(σ⊔**SetFunctionsOfOrder**◁σ▷[n])^σ⊔**SetFunctionsOfOrder**◁σ▷[n]

Unique isomorphism for **SetFunctions**◁ℕ▷ can be extended Sets◁ℕ▷ because every set in **Sets**◁ℕ▷ is defined by a characteristic function of **SetFunctions**◁ℕ▷ (see above):

Classical Direct Logic is much stronger than first-order axiomatizations of set theory.[87] Also, the semantics of Classical Direct Logic cannot be characterized using Tarskian Set Models [Tarski and Vaught 1957].[i]

**Theorem (Set Theory Model Soundness):** $\vdash_{\mathbf{Sets}\lhd\mathbb{N}\rhd}\Psi$ implies $\models_{\mathbb{S}\mathrm{ets}\lhd\mathbb{N}\rhd}\Psi$

Proof: Suppose $\vdash_{\mathbf{Sets}\lhd\mathbb{N}\rhd}\Psi$. The conclusion immediately follows because the axioms for the theory **Sets**◁ℕ▷ hold in the model 𝕊𝕖𝕥𝕤◁ℕ▷.

## Appendix 3. Historical development of inferential undecidability ("incompleteness")

### Truth versus Argumentation

Principia Mathematica [Russell 1925] (denoted by the theory Russell ) was intended to be a foundation for all of mathematics including Sets and Analysis building on [Frege 1879] that developed to characterizes the integers up to isomorphism [Peano 1889] as well as characterizing the real numbers up to isomorphism [Dedekind 1888] with the following theorems:

- *Full Peano Integers:* Let **X** be the structure <X, $0_X$, $S_X$>, then Peano[**X**] ⇒ **X**≈<ℕ, 0, S>[88] The theory Peano is the full theory of natural numbers with general induction that is strictly more powerful than cut-down first-order theory. Perhaps of greater import, there are nondeterministic Turing machines that Peano proves always halt that cannot be proved to halt in the cut-down first-order theory.

[i] See section on "Inadequacy of Tarskian Set Models."

- *Full Dedekind Reals:* Let **X** be the structure $<X, \leqq_X, 0_X, 1_X, +_X, *_X>$, then Dedekind[**X**] ⇒ $\mathbf{X} \approx <\mathbb{R}^{i}, \leqq, 0, 1, +, *>$[89]
  The theory Dedekind is the full theory of real numbers that is strictly more powerful than cut-down first-order theory.[90]

The above results categorically characterize the natural numbers (integers) and the real numbers up to isomorphism based on *argumentation*. There is no way to go beyond argumentation to get at some special added insight called "*truth*." Argumentation is all that we have.

Principia Mathematica [Russell 1903, 1925] was taken to formalize all of mathematics including numbers, points, manifolds, groups, *etc.* along with sets of these of these objects. Presumably metamathematics should follow suit and be formalized in Russell.

**Turing versus Gödel**

> *You shall not cease from exploration*
> *And the end of all our journeying*
> *Will be to arrive where we started*
> *And know the place for the first time.*
> T.S. Eliot [1942]

Turing recognized that proving that inference in mathematics is computationally undecidable is quite different than proving that there is a proposition of mathematics that is inferentially undecidable.[ii] [Turing 1936, page 259]:

> It should perhaps be remarked what I shall prove is quite different from the well-known results of Gödel [1931]. Gödel has shown that (in the formalism of Principia Mathematica) there are propositions U such that neither U nor ¬U is provable. … On the other hand, I shall show that there is no general method which tells whether a given formula U is provable.[91]

Although they share some similar underlying ideas, the method of proving computational undecidability developed by Church and Turing is much more robust than the one previously developed by Gödel that relies on "self-referential" propositions.

[i] ℝ is the set of real numbers
[ii] sometimes called "incompleteness."

The difference can be explicated as follows:

- Actors: an Actor that has an address for itself can be used to generate infinite computations.
- Propositions: "self-referential" propositions can be used to infer inconsistencies in mathematics.

As Wittgenstein pointed out, the following "self-referential" proposition leads an inconsistency in the foundations of mathematics: This proposition is not provable. If the inconsistencies of "self-referential" propositions stopped with this example, then it would be somewhat tolerable for an inconsistency-robust theory. However, other "self-referential" propositions (constructed in a similar way) can be used to prove every proposition thereby rendering inference useless.

This is why Direct Logic does not support "self-referential" propositions.[i]

## Contra Gödel et. al

The proof of the consistency of mathematics in this article contradicts the result [Gödel 1931] using "self-referential" propositions that mathematics cannot prove its own consistency.

One resolution is not to have "self-referential" propositions, which is contra Gödel *et. al.* Direct Logic aims to not have "self-referential" propositions by carefully arranging the rules so that "self-referential" propositions cannot be constructed. The basic idea is to use typed functions [Russell 1908, Church 1940] to construct propositions so that fixed points do not exist and consequently cannot be used to construct "self-referential" propositions.

## How the self-proof of consistency of mathematics was overlooked and then discovered

Before the paradoxes were discovered, not much attention was paid to proving consistency. Hilbert et. al. undertook to find a *convincing* proof of consistency. Gentzen found a consistency proof for the first-order Peano theory but many did not find it convincing because the proof was not elementary. Then following Carnap and Gödel, philosophers blindly accepted the necessity of "self-referential" prepositions in mathematics. And none of them seemed to understand Wittgenstein's critique. (Gödel insinuated that Wittgenstein was "*crazy.*")[ii] Instead, philosophers turned their attention to

---

[i] There It seems that are no practical uses for "self-referential" propositions in the mathematical foundations of Computer Science.

[ii] [Gödel 1931] proved the incompleteness results for Principia Mathematica as the foundation of all of mathematics. In opposition to Wittgenstein's devastating

exploring the question of which is the weakest theory in which Gödel's proof can be carried out. They were prisoners of the existing paradigm.

Computer scientists brought different concerns and a new perspective. They wanted foundations with the following characteristics:

- powerful so that arguments (proofs) are short and understandable and all logical inferences can be formalized
- standard so they can join forces and develop common techniques and technology
- inconsistency robust because computers deal in pervasively inconsistent information.

The results of [Gödel 1931], [Curry 1941], and [Löb 1055] played an important role the development of Direct Logic:

- Direct Logic easily formalized Wittgenstein's proof that Gödel's "self-referential" proposition leads to contradiction. So the consistency of mathematics had to be rescued against Gödel's "self-referential" proposition. The "self-referential" propositions used in results of [Curry 1941] and [Löb 1955] led to inconsistency in mathematics. So the consistency of mathematics had to be rescued against these "self-referential" propositions as well.
- Direct Logic easily proves the consistency of mathematics. So the consistency of mathematics had to be rescued against Gödel's "2nd incompleteness theorem."
- Direct Logic easily proves Church's Paradox. So the consistency of mathematics had to be rescued against the assumption that the theorems of mathematics can be computationally enumerated.

In summary, computer science advanced to a point where it caused the development of Direct Logic.

---

argument that "self-referential" propositions lead to contradictions in mathematics, Gödel later claimed that the results were for a the cut-down theory of first-order Peano numbers. In point of fact, any computationally undecidable mathematical theory in Direct Logic is inferentially undecidable.

## Inconsistency-robust Logic Programs

Logic Programs[i] can logically infer computational steps.

***Forward Chaining***

Forward chaining is performed using ⊢

(⊢$_{aTheory}$ *PropositionExpression*):*Continuation*
Assert *PropositionExpression* for *aTheory*.

(**When** ⊢$_{aTheory}$ *PropositionPattern* →
*Expression*):*Continuation*
When *PropositionPattern* holds for *aTheory*, evaluate *Expression*.

Illustration of forward chaining:

⊢$_t$ Human[Socrates]▮

**When** ⊢$_t$ Human[$x$] → ⊢$_t$ Mortal[$x$]▮

will result in asserting Mortal[Socrates] for theory t

***Backward Chaining***

Backward chaining is performed using ⊩

(⊩$_{aTheory}$ *GoalPattern* → *Expression*):*Continuation*
Set *GoalPattern* for *Theory* and when established evaluate *Expression*.

(⊩$_{aTheory}$ *GoalPattern*):*Expression*
Set *GoalPattern* for *Theory* and return a list of assertions that satisfy the goal.

(**When** ⊩$_{aTheory}$ *GoalPattern* → *Expression*):*Continuation*
When there is a goal that matches *GoalPattern* for Theory, evaluate *Expression*.

---

[i] [Church 1932; McCarthy 1963; Hewitt 1969, 1971, 2010; Milner 1972, Hayes 1973; Kowalski 1973]. Note that this definition of Logic Programs does *not* follow the proposal in [Kowalski 1973, 2011] that Logic Programs be restricted only to backward chaining, *e.g.,* to the exclusion of forward chaining, *etc.*

Illustration of backward chaining:

$\vdash_t$ Human[Socrates]▮

**When** $\Vdash_t$ Mortal[$x$] ⟶ ($\Vdash_t$ Human[$x$] ⟶ $\vdash_t$ Mortal[$x$])▮

$\Vdash_t$ Mortal[Socrates]▮

will result in asserting Mortal[Socrates] for theory t.

***SubArguments***

This section explains how subarguments[i] can be implemented in natural deduction.

**When** $\Vdash_s$ (*psi* $\vdash_t$ *phi*) →
  **Let** t' ← extension(t),
    **Do** $\vdash_{t'}$ *psi*,
      $\Vdash_{t'}$ *phi* → $\vdash_s$ (*psi* $\vdash_t$ *phi*))▮

Note that the following hold for t' because it is an extension of t:

- **When** $\vdash_t$ *theta* ⟶ $\vdash_{t'}$ *theta*▮
- **When** $\Vdash_{t'}$ *theta* ⟶ $\Vdash_t$ *theta*▮

[i] See appendix on Inconsistency Robust Natural Deduction.

**Index**

## End Notes

[1] Inference is direct when it does not involved unnecessary circumlocutions, *e.g.,* coding sentences as Godel numbers. In Direct Logic, it is possible speak directly about inference relationships.

[2] This section shares history with [Hewitt 2010b]

[3] D'Ariano and Tosini [2010] showed how the Minkowskian space-time emerges from a topologically homogeneous causal network, presenting a simple analytical derivation of the Lorentz transformations, with metric as pure event-counting.

> *Do events happen in space-time or is space-time that is made up of events? This question may be considered a "which came first, the chicken or the egg?" dilemma, but the answer may contain the solution of the main problem of contemporary physics: the reconciliation of quantum theory (QT) with general relativity (GR).Why? Because "events" are central to QT and "space-time" is central to GR. Therefore, the question practically means: which comes first, QT or GR? In spite of the evidence of the first position—"events happen in space-time"—the second standpoint—"space- time is made up of events"—is more concrete, if we believe à la Copenhagen that whatever is not "measured" is only in our imagination: space-time too must be measured, and measurements are always made-up of events. Thus QT comes first. How? Space-time emerges from the tapestry of events that are connected by quantum interactions, as in a huge quantum computer: this is the Wheeler's "It from bit."* [Wheeler 1990].

[4] According to [Law 2006], a classical realism (to which he does *not* subscribe) is:

> *Scientific experiments make no sense if there is no reality independent of the actions of scientists: an independent reality is one of conditions of possibility for experimentation. The job of the investigator is to experiment in order to make and test hypotheses about the mechanisms that underlie or make up reality. Since science is conducted within specific social and cultural circumstances, the models and metaphors used to generate fallible claims are, of course, socially contexted, and always revisable…Different 'paradigms' relate to (possibly different parts of) the same world.*

[5] Vardi [2010] has defended the traditional paradigm of proving that program meet specifications and attacked an early critical analysis as follows: "*With*

*hindsight of 30 years, it seems that De Millo, Lipton, and Perlis'* [1979] *article has proven to be rather misguid*ed." However, contrary to Vardi, limitations of the traditional paradigm of proving that program meet specifications have become much more apparent in the last 30 years—as admitted even by some who had been the most prominent proponents, *e.g.*, [Hoare 2003, 2009].

[6] According to [Hoare 2009]: *One thing I got spectacularly wrong. I could see that programs were getting larger, and I thought that testing would be an increasingly ineffective way of removing errors from them. I did not realize that the success of tests is that they test the programmer, not the program. Rigorous testing regimes rapidly persuade error-prone programmers (like me) to remove themselves from the profession. Failure in test immediately punishes any lapse in programming concentration, and (just as important) the failure count enables implementers to resist management pressure for premature delivery of unreliable code. The experience, judgment, and intuition of programmers who have survived the rigors of testing are what make programs of the present day useful, efficient, and (nearly) correct*.

[7] According to [Hoare 2009]: *Verification* [proving that programs meet specifications] *technology can only work against errors that have been accurately specified, with as much accuracy and attention to detail as all other aspects of the programming task. There will always be a limit at which the engineer judges that the cost of such specification is greater than the benefit that could be obtained from it; and that testing will be adequate for the purpose, and cheaper. Finally, verification* [proving that programs meet specifications] *cannot protect against errors in the specification itself.*

[8] Popper [1934] section 30.

[9] The thinking in almost all scientific and engineering work has been that models (also called theories or microtheories) should be internally consistent, although they could be inconsistent with each other.

Indeed some researchers have even gone so far as to construct consistency proofs for some small software systems, *e.g.*, [Davis and Morgenstern 2005] in their system for deriving plausible conclusions using classical logical inference for Multi-Agent Systems. In order to carry out the consistency proof of their system, Davis and Morgenstern make some simplifying assumptions:

- No two agents can simultaneously make a choice (following [Reiter 2001]).
- No two agents can simultaneously send each other inconsistent information.
- Each agent is individually serial, *i.e.,* each agent can execute only one primitive action at a time.
- There is a global clock time.

* Agents use classical Speech Acts (see [Hewitt 2006b 2007a, 2007c, 2008c]).
* Knowledge is expressed in first-order logic.

*The above assumptions are not particularly good ones for modern systems (e.g., using Web Services and many-core computer architectures).* [Hewitt 2007a]

The following conclusions can be drawn for documentation, use cases, and code of large software systems for human-computer interaction:

* Consistency proofs are impossible for whole systems.
* There are some consistent subtheories but they are typically mathematical. There are some other consistent microtheories as well, but they are small, make simplistic assumptions, and typically are inconsistent with other such microtheories [Addanki, Cremonini and Penberthy 1989].

Nevertheless, the Davis and Morgenstern research programme to prove consistency of microtheories can be valuable for the theories to which it can be applied. Also some of the techniques that they have developed may be able to be used to prove the consistency of the mathematical fragment of Direct Logic and to prove inconsistency robustness (see below in this article).

[10] Turing differed fundamentally on the question of inconsistency from Wittgenstein when he attended Wittgenstein's seminar on the Foundations of Mathematics [Diamond 1976]:

*Wittgenstein:... Think of the case of the Liar. It is very queer in a way that this should have puzzled anyone — much more extraordinary than you might think... Because the thing works like this: if a man says 'I am lying' we say that it follows that he is not lying, from which it follows that he is lying and so on. Well, so what? You can go on like that until you are black in the face. Why not? It doesn't matter. ...it is just a useless language-game, and why should anyone be excited?*

*Turing: What puzzles one is that one usually uses a contradiction as a criterion for having done something wrong. But in this case one cannot find anything done wrong.*

*Wittgenstein: Yes — and more: nothing has been done wrong, ... where will the harm come?*

*Turing: The real harm will not come in unless there is an application, in which a bridge may fall down or something of that sort.... You cannot be confident about applying your calculus until you know that there are no hidden contradictions in it.... Although you do not know that the bridge will fall if there are no contradictions, yet it is almost certain that if there are contradictions it will go wrong somewhere.*

Wittgenstein followed this up with [Wittgenstein 1956, pp. 104e–106e]: *Can we say: 'Contradiction is harmless if it can be sealed off'? But what prevents us from sealing it off?.*

[11] A more conservative axiomatization in Direct Logic is the following:

$\text{Policy}_1[x] \equiv \text{Sane}[x] \vdash_{Catch22} \text{Obligated}[x, \text{Fly}]$

$\text{Policy}_2[x] \equiv \text{Obligated}[x, \text{Fly}] \vdash_{Catch22} \text{Fly}[x]$

$\text{Policy}_3[x] \equiv \text{Crazy}[x] \vdash_{Catch22} \neg\text{Obligated}[x, \text{Fly}]$

$\text{Observe}_1[x] \equiv \neg\text{Obligated}[x, \text{Fly}].\neg\text{Fly}[x] \vdash_{Catch22} \text{Sane}[x]$

$\text{Observe}_2[x] \equiv \text{Fly}[x] \vdash_{Catch22} \text{Crazy}[x]$

$\text{Observe}_3[x] \equiv \text{Sane}[x], \neg\text{Obligated}[x, \text{Fly}] \vdash_{Catch22} \neg\text{Fly}[x]]$

$\text{Observe}_4 \equiv \vdash_{Catch22} \text{Sane}[\text{Yossarian}]$

$\text{Background}_2 \equiv \vdash_{Catch22} \neg\text{Obligated}[\text{Moon}, \text{Fly}]$

For the more conservative axiomatization above:

$\vdash_{Catch22} \text{Fly}[\text{Yossarian}]$ **but** $\vdash_{Catch22} \neg\text{Fly}[\text{Yossarian}]$

$\vdash_{Catch22} \neg\text{Fly}[\text{Yossarian}]$ **but** $\vdash_{Catch22} \text{Fly}[\text{Yossarian}]$

But, unlike for the stronger axiomatization using strong implication:

$\nvdash_{Catch22} \neg\text{Obligated}[\text{Yossarian}, \text{Fly}]$

$\nvdash_{Catch22} \neg\text{Sane}[\text{Yossarian}]$

[12] Because of the use of a very strong form of implication in the axiomatization, the following can also be inferred:

$\vdash_{Catch22} \neg\text{Obligated}[\text{Yossarian}, \text{Fly}]$

$\vdash_{Catch22} \neg\text{Sane}[\text{Yossarian}]$

[13] Philosophers have given the name *a priori* and *a posteriori* to the inconsistency

[14] including entanglement

[15] One possible approach towards developing inconsistency robust probabilities is to attach directionality to the calculations as follows:

**P1.** $\vdash_{Catch22} \mathbb{P}\text{Sane}[x] \overset{\leq}{\rightarrow} \mathbb{P}\text{Obligated}[x, \text{Fly}]$

**P2.** $\vdash_{Catch22} \mathbb{P}\text{Obligated}[x, \text{Fly}] \overset{\leq}{\rightarrow} \mathbb{P}\text{Fly}[x]$

**P3.** $\vdash_{Catch22} \mathbb{P}\text{Crazy}[x] \overset{\leq}{\rightarrow} \mathbb{P}\neg\text{Obligated}[x, \text{Fly}]$

**S1.** $\vdash_{Catch22} \mathbb{P}\neg\text{Obligated}[x, \text{Fly}] \wedge \neg\text{Fly}[x] \overset{\leq}{\rightarrow} \mathbb{P}\text{Sane}[x]$

**S2.** $\vdash_{Catch22} \mathbb{P}\text{Fly}[x] \overset{\leq}{\rightarrow} \mathbb{P}\text{Crazy}[x]$

**S3.** $\vdash_{Catch22} \mathbb{P}\text{Sane}[x] \wedge \neg\text{Obligated}[x, \text{Fly}] \overset{\leq}{\rightarrow} \mathbb{P}\neg\text{Fly}[x]$

**S4.** $\vdash_{Catch22} \mathbb{P}\text{Sane}[\text{Yossarian}] \overset{\sim}{\Rightarrow} 1$

Consequently, the following inferences hold

**I1.** $\vdash_{Catch22}$ ℙObligated[Yossarian, Fly] ⇝ 1  ⓘ **P1** *and* **S4**

**I2.** $\vdash_{Catch22}$ ℙFly[Yossarian] ⇝ 1  ⓘ *using* **P2** and **I1**

**I3.** $\vdash_{Catch22}$ ℙCrazy[Yossarian] ⇝1  ⓘ *using* **S2** *and* **I2**

**I4.** $\vdash_{Catch22}$ ℙ¬Obligated[Yossarian, Fly] ⇝ 1  ⓘ **P3** *and* **I3**

**I5.** $\vdash_{Catch22}$ ℙ¬Fly[Yossarian] ⇝ 0  ⓘ **I4** *and* **S3**

**I6.** $\vdash_{Catch22}$ ℙFly[Yossarian] ⇝ 1  ⓘ *reformulation of* **I5**

Thus there is a contradiction in *Catch22* in that both of the following hold in the above:

**I2.** $\vdash_{Catch22}$ **ℙFly[Yossarian] ⇝ 1**

**I6.** $\vdash_{Catch22}$ **ℙFly[Yossarian] ⇝ 0**

However, it is not possible to immediately conclude that **1≈0** because of the directionality.

[16] In [Law 2006]. Emphases added.

[17] In Latin, the principle is called *ex falso quodlibet* which means that from falsity anything follows.

[18] [Nekham 1200, pp. 288-289]; later rediscovered and published in [Lewis and Langford 1932]

[19] [Pospesel 2000] has discussed extraneous ∨ introduction on in terms of the following principle: Ψ, (Ψ∨Φ ⊦ Θ) ⊦ Θ

However, the above principle immediately derives extraneous ∨ introduction when Θ is Ψ∨Φ. In Direct Logic, argumentation of the above form would often be reformulated as follows to eliminate the spurious Φ middle proposition: Ψ, (Ψ ⊦ Θ) ⊦ Θ

[20] Direct Logic is distinct from the Direct Predicate Calculus [Ketonen and Weyhrauch 1984].

[21] The importance of (counter) examples in reasoning was emphasized in [Rissland 1984] citing mathematics, law, linguistics and computer science. According to [Gordon 2009]:

> *[Toulmin 1958] was one of the first to reflect on the limitations of mathematical logic as a model of rationality in the context of everyday discourse and practical problems. By the 1950s, logic had become more or less synonymous with mathematical logic, as invented by Boole, De Morgan, Pierce, Frege, Hilbert and others, starting in the middle of the nineteenth century. Interestingly, Toulmin proposed legal argumentation as a model for practical reasoning, claiming that normative models of practical reasoning should be measured by the ideals of jurisprudence. [Walton 2006] is a good starting point for getting an overview of the modern philosophy of argumentation.*

[22] in Rebecca Herold *Managing an Information Security and Privacy Awareness and Training Program* 2005. p. 101.

[23] although there is no claim concerning Euclid's own orientation

[24] *Cf.* "*on the ordinary notion of proof, it is compelling just because, presented with it, we cannot resist the passage from premises to conclusion without being unfaithful to the meanings we have already given to the expressions employed in it.*" [Dummett 1973]

[25] Rosemary Redfield. *Arsenic associated bacteria (NASA's claims)* RR Research blog. Dec. 6, 2010.

[26] Felisa Wolfe-Simon, et. al. *A bacterium that can grow by using arsenic instead of phosphorus* Science. Dec. 2, 2010.

[27] $$\text{Consequence}_1 \equiv \text{NaturalDeduction}(\text{Axiom}_2)$$
$$= \vdash_{\mathit{Achilles}} (A, B \vdash_{\mathit{Achilles}} Z)$$
$$\text{Consequence}_2 \equiv \text{Combination}(\text{Axiom}_1, \text{Consequence}_1)$$
$$= \vdash_{\mathit{Achilles}} A, B, (A, B \vdash_{\mathit{Achilles}} Z)$$
$$\text{Consequence}_3 \equiv \text{ForwardChaining}(\text{Consequence}_2)$$
$$= \vdash_{\mathit{Achilles}} Z$$
$$\text{ProofOfZ}[a_1, a_2] \equiv$$
$$\text{ForwardChaining}[\text{Combination}[a_1,\ \text{NaturalDeduction}[a_2]]]$$

[28] McGee [1985] has challenged modus ponens using an example that can be most simply formalized in Direct Logic as follows:

$$\text{RepublicanWillWin} \vdash_{\mathit{McGee}} (\neg\text{ReaganWillWin} \vdash_{\mathit{McGee}} \text{AndersonWillWin})$$

and $\vdash_{\mathit{McGee}} \text{RepublicanWillWin}$

From the above, in Direct Logic it follows that:

$$\neg\text{ReaganWillWin} \vdash_{\mathit{McGee}} \text{AndersonWillWin}$$

McGee challenged the reasonableness of the above conclusion on the grounds that. intuitively, the proper inference is that if Reagan will not win, then $\neg$AndersonWillWin because Carter (the Democratic candidate) will win. However, in theory *McGee*, it is reasonable to infer AndersonWillWin from $\neg$ReaganWillWin because RepublicanWillWin holds in *McGee*.

McGee phrased his argument in terms of implication which in Direct Logic (see following discussion in this paper) would be as follows:

$$\vdash_{\mathit{McGee}} \text{RepublicanWillWin} \Rightarrow (\neg\text{ReaganWillWin} \Rightarrow \text{AndersonWillWin})$$

However, this makes no essential difference because, in Direct Logic, it still follows that $\vdash_{\mathit{McGee}} (\neg\text{ReaganWillWin} \Rightarrow \text{AndersonWillWin})$

[29] [*cf.* Church 1934, Kleene 1936]

[30] Direct inference is defined differently in this paper from probability theory [Levy 1977, Kyburg and Teng 2001], which refers to "*direct inference*" of

frequency in a reference class (the most specific class with suitable frequency knowledge) from which other probabilities are derived.

[31] [Jaśkowski 1934][31] that doesn't require artifices such as indices (labels) on propositions or restrictions on reiteration

[32] This section of the paper shares some history with [Hewitt 2010b].

[33] Turing [1936] stated:

- *the behavior of the computer at any moment is determined by the symbols which he* [the computer] *is observing, and his 'state of mind' at that moment*
- *there is a bound B to the number of symbols or squares which the computer can observe at one moment. If he wishes to observe more, he must use successive observation*s.

Gödel's conception of computation was formally the same as Turing but more reductionist in motivation:

*There is a major difference between the historical contexts in which Turing and Gödel worked. Turing tackled the Entscheidungsproblem* [computational decidability of provability] *as an interesting mathematical problem worth solving; he was hardly aware of the fierce foundational debates. Gödel on the other hand, was passionately interested in the foundations of mathematics. Though not a student of Hilbert, his work was nonetheless deeply entrenched in the framework of Hilbert's finitistic program, whose main goal was to provide a meta-theoretic finitary proof of the consistency of a formal system "containing a certain amount of finitary number theory."* Shagrir [2006]

[34] According to [Turing 1948]:

*LCMs* [Logical Computing Machines: Turing's expression for Turing machines] *can do anything that could be described as ... "purely mechanical"...This is sufficiently well established that it is now agreed amongst logicians that "calculable by means of an LCM" is the correct accurate rendering* [of phrases like "*purely mechanical*"]

[35] [Wang 1974, p. 84]

[36] An example of the global state model is the Abstract State Machine (ASM) model [Blass, Gurevich, Rosenzweig, and Rossman 2007a, 2007b; Glausch and Reisig 2006].

[37] This result is very old. It was known by Dijkstra motivating his belief that it is impossible to implement unbounded nondeterminism. Also the result played a crucial role in the invention of the Actor Model in 1972.

Consider the following Nondeterministic Turing Machine:

*Step 1*: Next do either *Step 2* or *Step 3.*
*Step 2*: Next do *Step 1.*
*Step 3*: Halt.

It is possible that the above program does not halt. It is also possible that the above program halts.

Note that above program is not equivalent to the one below in which it is not possible to halt:

*Step 1*: Next do *Step 1*.

[38] This proof does not apply to extensions of Nondeterministic Turing Machines that are provided with a new primitive instruction NoLargest which is defined to write an unbounded large number on the tape. Since executing NoLargest can write an unbounded amount of tape in a single instruction, executing it can take an unbounded time during which the machine cannot read input.

Also, the NoLargest primitive is of limited practical use. Consider a Nondeterministic Turing Machine with two input-only tapes that can be read nondeterministically and one standard working tape.

It is possible for the following program to copy both of its input tapes onto its working tape:

*Step 1: Either*

1. copy the current input from the 1st input tape onto the working tape and next do *Step 2,*

*or*

2. copy the current input from the 2nd input tape onto the working tape and next do *Step 3.*

*Step 2:* Next do *Step 1*.

*Step 3:* Next do *Step 1*.

It is also possible that the above program does not read any input from the 1st input tape (*cf.* [Knabe 1993]) and the use of NoLargest is of no use in alleviating this problem. Bounded nondeterminism is a symptom of deeper underlying issues with Nondeterministic Turing Machines.

[39] Consequently,

- The tree has an infinite path. ⇔ The tree is infinite. ⇔ It is possible that P does not halt. If it is possible that P does not halt, then it is possible that that the set of outputs with which P halts is infinite.
- The tree does not have an infinite path. ⇔ The tree is finite. ⇔ P always halts. If P always halts, then the tree is finite and the set of outputs with which P halts is finite.

[40] Arbiters render meaningless the states in the Abstract State Machine (ASM) model [Blass, Gurevich, Rosenzweig, and Rossman 2007a, 2007b; Glausch and Reisig 2006].

[41] The logic gates require suitable thresholds and other characteristics.

[42] *cf.* denotational semantics of the lambda calculus [Scott 1976]

[43] Proof: Suppose to obtain a contraction that ComputationallyDecidable[HaltingProblem].

Define the program Diagonal as follows:

Diagonal ≡ [x]→ Halt▪[x, x] � **True** ⦂ InfiniteLoop▪[ ], **False** ⦂ **True**

where InfiniteLoop ≡ [ ]→ InfiniteLoop▪[ ]

Poof of inconsistency: By the definition of Diagonal:

⌊Diagonal⌋▪[Diagonal] →₁ Halt▪[Diagonal, Diagonal] �
True ⦂ InfiniteLoop▪[ ],
False ⦂ True

Consider the following 2 cases:

1. Halt▪[Diagonal, Diagonal] →₁ *True*
   Converges[⌊Diagonal⌋▪[Diagonal]] by the axioms for Halt
   ¬Converges[⌊Diagonal⌋▪[Diagonal]] by the definition of Diagonal
2. Halt▪[Diagonal, Diagonal] →₁ *False*
   ¬Converges[⌊Diagonal⌋▪[Diagonal]] by the axioms for Halt
   Converges[⌊Diagonal⌋▪[Diagonal]] by the definition of Diagonal

Consequently, ¬ComputationallyDecidable[HaltingProblem]

[44] Note that this theorem is very different from the result [Kleene 1938], that mathematics can be extended with a proposition asserting its own consistency.

[45] A prominent logician referee of this article suggested that if the proof is accepted then consistency should be made an explicit premise of every theorem of classical mathematics!

[46] As shown above, there is a simple proof in Classical Direct Logic that Mathematics (⊢) is consistent. If Classical Direct Logic has a bug, then there might also be a proof that Mathematics is inconsistent. Of course, if a such a bug is found, then it must be repaired.

Fortunately, Classical Direct Logic is simple in the sense that it has just *one* fundamental axiom:

∀[P:**Boolean**^ℕ]→ Inductive[P]⇨ ∀[i:ℕ]→ P[i]
*where* ∀[P:**Boolean**^ℕ]→
Inductive[P] ⇔ (P[0] ∧ ∀[i:ℕ]→ P[i] ⇨P[i+1])

Of course, Classical Direct Logic has machinery in addition the above axiom that could also have bugs.

The Classical Direct Logic proof that Mathematics (⊢) is consistent is very robust. One explanation is that consistency is built in to the very architecture of classical mathematics because it was designed to be consistent. Consequently, it is not absurd that there is a simple proof of the consistency of Mathematics (⊢) that does not use all of the machinery of Classical Direct Logic.

In reaction to paradoxes, philosophers developed the dogma of the necessity of strict separation of “object theories” (theories about basic mathematical entities such as numbers) and “meta theories” (theories about theories). This linguistic separation can be very awkward in Computer Science. Consequently, Direct Logic does not have the separation in order that some propositions can be more “directly” expressed. For example, Direct Logic can use ⊢⊢Ψ to express that it is provable that P is provable in Mathematics. It turns out in Classical Direct Logic that $\vdash\vdash\Psi$ holds if and only if $\vdash\Psi$ holds. By using such expressions, Direct Logic contravenes the philosophical dogma that the proposition $\vdash\vdash\Psi$ must be expressed using Gödel numbers.

[47] As shown above, there is a simple proof in Classical Direct Logic that Mathematics ($\vdash$) is consistent. If Classical Direct Logic has a bug, then there might also be a proof that Mathematics is inconsistent. Of course, if a such a bug is found, then it must be repaired.

Fortunately, Classical Direct Logic is simple in the sense that it has one fundamental axiom:

$$\forall[P{:}\mathbf{Boolean}^{\mathbb{N}}]\rightarrow \text{Inductive}[P] \Rightarrow \forall[i{:}\mathbb{N}]\rightarrow P[i]$$

$$\textit{where}\ \forall[P{:}\mathbf{Boolean}^{\mathbb{N}}]\rightarrow$$

$$\text{Inductive}[P] \Leftrightarrow P[0] \wedge \forall[i{:}\mathbb{N}]\rightarrow P[i] \Rightarrow P[i+1]$$

Of course, Classical Direct Logic has machinery in addition the above axiom that could also have bugs.

The Classical Direct Logic proof that Mathematics ($\vdash$) is consistent is very robust. One explanation is that consistency is built in to the very architecture of classical mathematics because it was designed to be consistent. Consequently, it is not absurd that there is a simple proof of the consistency of Mathematics ($\vdash$) that does not use all of the machinery of Classical Direct Logic.

In reaction to paradoxes, philosophers developed the dogma of the necessity of strict separation of "object theories" (theories about basic mathematical entities such as numbers) and "meta theories" (theories about theories). This linguistic separation can be very awkward in Computer Science. Consequently, Direct Logic does not have the separation in order that some propositions can be more "directly" expressed. For example, Direct Logic can use $\vdash\vdash\Psi$ to express that it is provable that P is provable in Mathematics. It turns out in Classical Direct Logic that $\vdash\vdash\Psi$ holds if and only if $\vdash\Psi$ holds. By using such expressions, Direct Logic contravenes the philosophical dogma that the proposition $\vdash\vdash\Psi$ must be expressed using Gödel numbers.

[48] This integration can include calendars and to-do lists, communications (including email, SMS, Twitter, Facebook), presence information (including who else is in the neighborhood), physical (including GPS recordings), psychological (including facial expression, heart rate, voice stress) and social (including family, friends, team mates, and colleagues), maps (including firms, points of interest, traffic, parking, and weather), events (including alerts and status), documents (including presentations, spreadsheets, proposals, job applications, health records, photons, videos, gift lists, memos, purchasing, contracts, articles), contacts (including social

graphs and reputation), purchasing information (including store purchases, web purchases, GPS and phone records, and buying and travel habits), government information (including licenses, taxes, and rulings), and search results (including rankings and rating).

[49] In 1994, Alan Robinson noted that he has "*always been a little quick to make adverse judgments about what I like to call 'wacko logics' especially in Australia…I conduct my affairs as though I believe … that there is only one logic. All the rest is variation in what you're reasoning about, not in how you're reasoning …* [Logic] *is immutable*." (quoted in Mackenzie [2001] page 286)

On the other hand Richard Routley noted:

> *… classical logic bears a large measure of responsibility for the growing separation between philosophy and logic which there is today… If classical logic is a modern tool inadequate for its job, modern philosophers have shown a classically stoic resignation in the face of this inadequacy. They have behaved like people who, faced with a device, designed to lift stream water, but which is so badly designed that it spills most of its freight, do not set themselves to the design of a* better *model, but rather devote much of their energy to constructing ingenious arguments to convince themselves that the device is admirable, that they do not need or want the device to deliver more water; that there is nothing wrong with wasting water and that it may even be desirable; and that in order to "improve" the device they would have to change some features of the design, a thing which goes totally against their engineering intuitions and which they could not possibly consider doing.* [Routley 2003]

[50] According to [Kuhn 1962 page 151]

> *And Max Planck, surveying his own career in his Scientific Autobiography* [Planck 1949]*, sadly remarked that "a new scientific truth does not triumph by convincing its opponents and making them see the light, but rather because its opponents eventually die, and a new generation grows up that is familiar with it."*

[51] It is not possible to guarantee the consistency of information because consistency testing is computationally undecidable even in logics much weaker than first order logic. Because of this difficulty, it is impractical to test whether information is consistent.

[52] Consequently iDescriber makes use of direct inference in Direct Logic to reason more safely about inconsistent information because it omits the rules of classical logic that enable every proposition to be inferred from a single inconsistency.

[53] By the *Computational Representation Theorem* [Clinger 1981; Hewitt 2006], which can define all the possible executions of a procedure.\

[54] e.g. [Shulman 2012, nLab 2014]

[55] **ℕ** is the type of Natural Numbers.

[56] type of 2-element list with first element of type $\sigma_1$ and with second element of type $\sigma_2$

[57] type of computable procedures from type $\sigma_1$ into $\sigma_2$.

[58] type of functions from $\sigma 1$ into $\sigma 2$

[59] type of term of type $\sigma$

[60] *if* **t** *then* $\Phi_1$ *else* $\Phi_2$

[61] $\Phi_1$, … and $\Phi_k$ infer $\Psi_1$, …, and $\Psi_n$

[62] **p** is a proof of $\Phi$

[63] parameterized mutually recursive definitions of $v_{1\text{ to }n}\triangleleft\tau_{1\text{ to }n}\triangleright$

[64] *if* $t_1$ *then* $t_2$ *else* $t_3$

[65] Because there is no type restriction, fixed points may be freely used to define recursive procedures on expressions.

[66] *if* **t** *then* $s_1$ *else* $s_1$

[67] Proof: $(\Psi \vee (\Phi\wedge\Psi)) \Leftrightarrow (\Psi\vee \Phi) \wedge (\Psi\vee\Psi) \Leftrightarrow (\Psi\vee \Phi) \wedge \Psi$

[68] Direct Logic uses the full meaning of quantification as opposed to a cut down syntactic variant, e.g., [Henken 1950]. Disadvantages of the Henkin approach are explained in [Restall 2007].

[69] [Church 1956; Concoran 1973, 1980; Boulos 1975; Shapiro 2002]

[70] "*The world that appears to our senses is in some way defective and filled with error, but there is a more real and perfect realm, populated by entities* [called "ideals" or "forms"] *that are eternal, changeless, and in some sense paradigmatic for the structure and character of our world. Among the most important of these* [ideals] *(as they are now called, because they are not located in space or time) are Goodness, Beauty, Equality, Bigness, Likeness, Unity, Being, Sameness, Difference, Change, and Changelessness. (These terms — "Goodness", "Beauty", and so on — are often capitalized by those who write about Plato, in order to call attention to their exalted status;…) The most fundamental distinction in Plato's philosophy is between the many observable objects that appear beautiful (good, just, unified, equal, big) and the one object that is what Beauty (Goodness, Justice, Unity) really is, from which those many beautiful (good, just, unified, equal, big) things receive their names and their corresponding characteristics. Nearly every major work of Plato is, in some way, devoted to or dependent on this distinction.*

*Many of them explore the ethical and practical consequences of conceiving of reality in this bifurcated way. We are urged to transform our values by taking to heart the greater reality of the* [ideals] *and the defectiveness of the corporeal world."* [Kraut 2004]

[71] Structuralism takes a different view of mathematics:

*The structuralist vigorously rejects any sort of ontological independence among the natural numbers. The essence of a natural number is its relations to other natural numbers. The subject matter of arithmetic is a single abstract structure, the pattern common to any infinite collection of*

*objects that has a successor relation, a unique initial object, and satisfies the induction principle. The number 2 is no more and no less than the second position in the natural number structure; and 6 is the sixth position. Neither of them has any independence from the structure in which they are positions, and as positions in this structure, neither number is independent of the other.* [Shapiro 2000]

72 Basic axioms are as follows:

**True** � **True** ⦂ $E_1$, **False** ⦂ $E_2 \longrightarrow E_1$

**False** � **False** ⦂ $E_1$, **True** ⦂ $E_2 \longrightarrow E_1$

**False** � **True** ⦂ $E_1$, **False** ⦂ $E_2 \longrightarrow E_2$

**True** � **False** ⦂ $E_1$, **True** ⦂ $E_2 \longrightarrow E_2$

$(E_1 \longrightarrow E_2) \wedge (E_2 \longrightarrow E_3)) \Rightarrow (E_1 \longrightarrow E_3)$

$([x] \rightarrow F[x])[E] \longrightarrow F[E]$

($E_1$ **either** $E_2$) $\longrightarrow E_1$[72]

($E_1$ **either** $E_2$) $\longrightarrow E_2$[72]

$F_1 \longrightarrow F_2 \Rightarrow F_1(E) \longrightarrow F_2(E)$

ⓘ *an application evolves if its operator evolves*

$E_1 \longrightarrow E_2 \Rightarrow F(E_1) \longrightarrow F(E_2)$

ⓘ *an application evolves if its operand evolves*

$E_1 \longrightarrow E_2 \Rightarrow (\downarrow E_2 \Rightarrow \downarrow E_1)$

$E_1 \underline{\downarrow} E_2 \Leftrightarrow ((E_1 \longrightarrow E_2 \wedge \underline{\downarrow} E_2) \vee (\underline{\downarrow} E_1 \wedge E_1 = E_2))$

$E \downarrow_1 \Leftrightarrow (E \underline{\downarrow} \wedge (E \underline{\downarrow} E_1 \wedge E \underline{\downarrow} E_2) \Rightarrow E_1 = E_1)$

$\underline{\downarrow} E_1 \Rightarrow \neg (E_1 \longrightarrow E_2)$

73 [Church 1956; Boolos 1975; Corcoran 1973, 1980]

74 along with lists

75 Set◁σ▷ is the type of a set of type **σ**, Sets◁σ▷ is the type all sets of sets over type **σ**, and Domain◁σ▷=σ⊔Sets◁σ▷ with the following axioms:

{ }:**Set**◁σ▷ ⓘ the empty set { } is a set

∀[x:σ]→ {x}:**Set**◁σ▷ ⓘ a singleton set is a set

∀[s:**Sets**◁σ▷]→ ⋃s:**Sets**◁σ▷ ⓘ all elements of the subsets of a set is a set

∀[x:σ]→ x∉{ } ⓘ the empty set { } has no elements

∀[s:**Set**◁σ▷, f:$\sigma^{\sigma}$] → (Elementwise[f])[s]:Set◁σ▷

ⓘ the function image of a set is a set

∀[s:**Set**◁σ▷, p:**Boolean**$^{\sigma}$] → s↾p:**Set**◁σ▷

ⓘ a predicate restriction of a set is a set

∀[s:**Set**◁σ▷]→ { }⊆s ⓘ { } is a subset of every set

∀[$s_1$,$s_2$:**Set**◁σ▷]→ $s_1 = s_2 \Leftrightarrow (\forall [x:\sigma] \rightarrow x \in s_1 \Leftrightarrow x \in s_2)$

$\forall[x,y{:}\sigma]\rightarrow x\in\{y\} \Leftrightarrow x=y$

$\forall[s_1,s_2{:}\mathbf{Set}\lhd\sigma\rhd]\rightarrow s_1\subseteq s_2 \Leftrightarrow \forall[x{:}\sigma]\rightarrow x\in s_1 \Rightarrow x\in s_2$

$\forall[x{:}\sigma;\ s_1,s_2{:}\mathbf{Set}\lhd\sigma\rhd]\rightarrow x\in s_1\cup s_2 \Leftrightarrow (x\in s_1 \vee x\in s_2)$

$\forall[x{:}\sigma;\ s1,s2{:}\mathbf{Set}\lhd\sigma\rhd]\rightarrow x\in s1\cap s2 \Leftrightarrow (x\in s1 \wedge x\in s2)$

$\forall[x{:}\mathbf{Domain}\lhd\sigma\rhd;\ s{:}\mathbf{Sets}\lhd\sigma\rhd]\rightarrow x\in\bigcup s \Leftrightarrow \exists[s1{:}\mathrm{Sets}\lhd\sigma\rhd]\rightarrow x\in s1 \wedge s1\in s$

$\forall[y{:}\sigma;\ s{:}\mathbf{Set}\lhd\sigma\rhd,\ f{:}\sigma^{\sigma}]\rightarrow$

$y\in(\mathrm{Elementwise}[f])[s] \Leftrightarrow \exists[x\in s]\rightarrow f[x]=y$

$\forall[y{:}\sigma;\ s{:}\mathbf{Set}\lhd\sigma\rhd,\ p{:}\mathbf{Boolean}^{\sigma}]\rightarrow y\in s\upharpoonright p \Leftrightarrow y\in s \wedge p[y]$

The natural numbers are axiomatised as follows where Successor is the successor function:

- $0{:}\mathbb{N}$
- $\mathrm{Successor}{:}\mathbb{N}^{\mathbb{N}}$
- $\forall[i{:}\mathbb{N}]\rightarrow \mathrm{Successor}[i]\neq 0$
- $\forall[i, j{:}\mathbb{N}]\rightarrow \mathrm{Successor}[i]=\mathrm{Successor}[j] \Rightarrow i=j$
- $\forall[P{:}\mathbf{Boolean}^{\mathbb{N}}]\rightarrow \mathrm{Inductive}[P]\Rightarrow \forall[i{:}\mathbb{N}]\rightarrow P[i]$

  where

  $\forall[P{:}\mathbf{Boolean}^{\mathbb{N}}]\rightarrow \mathrm{Inductive}[P]$

  $\Leftrightarrow P[0] \wedge \forall[i{:}\mathbb{N}]\rightarrow P[i] \Rightarrow P[\mathrm{Successor}[i]]$

[76] *I.e.*, $\nexists[s{:}\mathbf{Sets}\lhd\mathbb{N}\rhd]\rightarrow \forall[e{:}\mathbf{Domain}\lhd\mathbb{N}\rhd]\rightarrow e\in s \Leftrightarrow e{:}\mathbf{Sets}\lhd\mathbb{N}\rhd$

where $\mathbf{Domain}\lhd\mathbb{N}\rhd=\mathbb{N}\sqcup\mathbf{Sets}\lhd\mathbb{N}\rhd$

[77] a set is not well founded if and only if it has an infinite $\in$ chain

[78] Quoted by Bob Boyer [personal communication 12 Jan. 2006].

[79] Atomics *and* Elements *are disjoint*

[80] For example, there is no restriction that an inductive predicate must be defined by a first order proposition.

[81] [Dedekind 1888], [Peano 1889], and [Zermelo 1930].

[82] [Dedekind 1888, Peano 1889]

[83] $\mathbb{N}$ is identified with the type of natural numbers

[84] [Dedekind 1888], [Peano 1889], and [Zermelo 1930].

[85] $\mathbb{R}$ is identified with the type of real numbers

[86] *cf.* [Zermelo 1930].

[87] The Continuum Hypothesis remains an open problem for Direct Logic because its set theory is very powerful. The forcing technique used to prove the independence of the Continuum Hypothesis for first-order set theory [Cohen 1963] does not apply to Direct Logic because of the strong induction axiom [Dedekind 1888, Peano 1889] used in formalizing the natural numbers $\mathbb{N}$, which is the foundation of set theory. Of course, trivially,

$(\vDash_{\mathbf{Domain}\lhd\mathbb{N}\rhd}\mathrm{ContinuumHypothesis})\vee(\vDash_{\mathbf{Domain}\lhd\mathbb{N}\rhd}\neg\mathrm{ContinuumHypothesis})$

where $\mathbf{Domain}\lhd\sigma\rhd=\sigma\sqcup\mathbf{Sets}\lhd\sigma\rhd$.

[88] Peano[X], means that **X** satisfies the full Peano axioms for the non-negative integers, ℕ is the type of non-negative integers, s is the successor function, and ≈ means isomorphism.

The isomorphism is proved by defining a function f from ℕ to X by:

1. $f[0]=0_X$
2. $f[S[n]]=S_X[f[n]]$

Using proof by induction, the following follow:

1. f is defined for every element of ℕ
2. f is one-to-one

Proof:

First prove $\forall[n\in X]\rightarrow f[n]=0_X \Rightarrow n=0$

*Base*: Trivial.

*Induction*: Suppose $f[n]=0_X \Rightarrow n=0$

$f[S[n]]=S_X[f[n]]$ Therefore if $f[S[n]]=0_X$ then $0_X=S_X[f[n]]$ which is an inconsistency

Suppose $f[n]=f[m]$. To prove: $n=m$

Proof: By induction on n:

*Base*: Suppose $f[0]=f[m]$. Then $f[m]= 0_X$ and $m=0$ by above

*Induction*: Suppose $\forall[m\in\mathbb{N}]\rightarrow f[n]=f[m]\Rightarrow n=m$

Proof: By induction on m:

*Base*: Suppose $f[n]=f[0]$. Then $n=m=0$

*Induction*: Suppose $f[n]=f[m]\Rightarrow n=m$

$f[S[n]]=S_X[f[n]]$ and $f[S[m]]=S_X[f[m]]$

Therefore $f[S[n]]=f[S[m]]\Rightarrow S[n]=S[m]$

3. the range of f is all of X.

Proof: To show: Inductive[Range[f]]

*Base*: To show $0_X\in$Range[f]. Clearly $f[0]=0_X$

*Induction*: To show $\forall[n\in \text{Range}[f]]\rightarrow S_X[n]\in \text{Range}[f]$.

Suppose that $n\in$Range[f]. Then there is some m such that $f[m]=n$.

To prove: $\forall[k\in\mathbb{N}]\rightarrow f[k]=n \Rightarrow S_X[n]\in \text{Range}[f]$

Proof: By induction on k:

*Base*: Suppose $f[0]=n$. Then $n= 0_X =f[0]$ and $S_X[n]=f[S[0]]\in \text{Range}[f]$

*Induction*: Suppose $f[k]=n \Rightarrow S_X[n]\in \text{Range}[f]$

Suppose $f[S[k]]=n$. Then $n=S_X[f[k]]$ and

$S_X[n]=S_X[S_X[f[k]]]=S_X[f[S[k]]]= f[S[S[k]]]\in \text{Range}[f]$

[89] Dedekind[X], means that **X** satisfies the Dedekind axioms for the real numbers

[90] Robinson [1961]

[91] The inferability problem is to computationally decide whether a proposition defined by sentence is inferable.

Theorem [Church 1935 followed by Turing 1936]:

$$\text{Consistent}_T \Rightarrow \neg\text{ComputationallyDecidable}[\text{InferenceProblem}_T]$$

Proof. Suppose to obtain a contradiction that $\text{ComputationallyDecidable}[\text{InferenceProblem}_T]$.

This means that there is a total computational deterministic predicate $\text{Inferable}_T$ such that the following 3 properties hold

1. $\text{Inferable}_{T\bullet}[\Psi] \twoheadrightarrow_1 \textbf{True} \Leftrightarrow \vdash_T \Psi$
2. $\text{Inferable}_{T\bullet}[\Psi] \twoheadrightarrow_1 \textbf{False} \Leftrightarrow \nvdash_T \Psi$
3. $\text{Inferable}_{T\bullet}[\Psi] \twoheadrightarrow_1 \textbf{True} \vee \text{Inferable}_{T\bullet}[\Psi] \twoheadrightarrow_1 \textbf{False}$

The proof proceeds by showing that if inference is computationally decidable, the halting problem is computationally decidable.

Consider proposition of the form $\text{Converges}[\lfloor p \rfloor_\bullet[x]]$ , which is the proposition that the program p halts on input x.

Lemma: $\text{Consistent}_T \Rightarrow \text{Inferable}_{T\bullet}[\text{Converges}[\lfloor p \rfloor_\bullet[x]]] \twoheadrightarrow_1 \textbf{True}$ if and only if $\text{Converges}[\lfloor p \rfloor_\bullet[x]]$

Proof of lemma: Suppose $\text{Consistent}_T$

1. Suppose $\text{Inferable}_{T\bullet}[\text{Converges}[\lfloor p \rfloor_\bullet[x]]] \twoheadrightarrow_1 \textbf{True}$. Then $\vdash_T \text{Converges}[\lfloor p \rfloor_\bullet[x]]$ by definition of $\text{Inferable}_T$. Suppose to obtain a contradiction that $\neg\text{Converges}[\lfloor p \rfloor_\bullet[x]]$. The contradiction $\nvdash_T \text{Converges}[\lfloor p \rfloor_\bullet[x]]$ follows by consistency of $T$.
2. Suppose $\text{Converges}[\lfloor p \rfloor_\bullet[x]]$. Then $\vdash_T \text{Converges}[\lfloor p \rfloor_\bullet[x]]$ by Adequacy of $T$. It follows that $\text{Inferable}_{T\bullet}[\text{Converges}[\lfloor p \rfloor_\bullet[x])\ ] \twoheadrightarrow_1 \textbf{True}$.

But this contradicts $\neg\text{ComputationallyDecidable}[\text{HaltingProblem}]$ because $\text{Halt}[p, x] \Leftrightarrow \text{Inferable}_{T\bullet}[\text{Converges}[\lfloor p \rfloor_\bullet[x]]]$

Consequently,

$$\text{Consistent}_T \Rightarrow \neg\text{ComputationallyDecidable}[\text{InferenceProblem}_T]$$